\documentclass[]{aa}
\usepackage{graphicx}
\usepackage{txfonts}
\usepackage{natbib}
\usepackage[]{hyperref}
\hypersetup{
    bookmarks=true,                                   
    pdftitle={}, 
    pdfauthor={}, 
    colorlinks=true,                                  
    linkcolor=blue,                                   
    citecolor=blue,                                   
    urlcolor=blue                                     
}

\newcommand{\msun}{M$_{\odot}$}

\newcommand{\nurot}{\nu_{\mathrm{rot}}}

\begin{document}

    \title{Deciphering the oscillation spectrum of $\gamma$ Doradus and SPB stars} 
    \subtitle{}

    \author{S. Christophe\inst{1,2} 
		\and 
		J. Ballot\inst{2} 
		\and 
		R.-M. Ouazzani\inst{1,3}
		\and
		V. Antoci\inst{3}
		\and
		S.J.A.J. Salmon\inst{4}}
    \institute{
		LESIA, Observatoire de Paris, PSL Research University, CNRS, Sorbonne Universit\'es, UPMC Univ. Paris 06, Univ. Paris Diderot,
		Sorbonne Paris Cit\'e, 5 place Jules Janssen, 92195 Meudon, France
		\and
		IRAP, Universit\'e de Toulouse, CNRS, UPS, CNES, 14 avenue \'Edouard Belin, 31400 Toulouse, France
           \and
		 Stellar Astrophysics Centre, Department of Physics and Astronomy, Aarhus University, Ny Munkegade 120, DK-8000 Aarhus C, 				Denmark
		\and
		STAR Institute, Universit\'e de Li\`ege, All\'ee du 6 Ao\^ut 19, 4000 Li\`ege, Belgium
               }
    \date{Draft: \today; Received xxx; accepted xxx}
    
    \abstract{The space-based  \emph{Kepler} mission provided four years of highly precise and almost uninterrupted photometry for hundreds of $\gamma$ Doradus stars and tens of SPB stars, finally allowing us to apply asteroseismology to these gravity mode pulsators. Without rotation, gravity modes are equally spaced in period. This simple structure does not hold in rotating stars for which rotation needs to be taken into account to accurately interpret the oscillation spectrum.
	}{We aim to develop a stellar-model-independent method to analyse and interpret the oscillation spectrum of $\gamma$ Dor and SPB stars.
	}{Within the traditional approximation of rotation, we highlight the possibility of recovering the equidistance of period spacings by stretching the pulsation periods. The stretching function depends on the degree and azimuthal order of gravity modes and the rotation rate of the star. In this new stretched space, the pulsation modes are regularly spaced by the stellar buoyancy radius.
	}{On the basis of this property, we implemented a method to search for these new regularities and simultaneously infer the rotation frequency and buoyancy radius. Tests on synthetic spectra computed with a non-perturbative approach show that we can retrieve  these two parameters with reasonable accuracy along with the mode identification. In uniformly rotating models of a typical $\gamma$ Dor star, and for the most observed prograde dipole modes, we show that the accuracy on the derived parameters is better than 5\% on both the internal rotation rate and the buoyancy radius. Finally, we apply the method to two stars of the \emph{Kepler} field, a $\gamma$ Dor and an SPB, and compare our results with those of other existing methods.
	}{We provide a stellar-model-independent method to obtain the near-core rotation rate, the buoyancy radius and mode identification from g-mode spectra of $\gamma$ Dor and SPB stars.}
             
    
    \titlerunning{Deciphering the oscillation spectrum of SPB and $\gamma$ Doradus stars}
	\authorrunning{Christophe et al.}
    
    \maketitle
	
%
%

\section{Introduction}
\label{sec:intro}

In stellar physics, the treatment of interfaces between convective and radiative regions remains rather simplistic and subject to large uncertainties despite the critical effects on the stellar structure and evolution. This is especially true for main-sequence stars that possess a convective core. Indeed, mixing processes at the interface can extend the central mixed zone and increase the amount of hydrogen fuel available for nuclear fusion reactions i.e. the time spent on the main sequence. A better treatment of these interfaces will thereby result in more precise determinations of stellar ages. In standard models, the convective core boundary is set by the Schwarzschild criterion that is equivalent to the location where the acceleration of convective motions cancels out. Comparison of standard stellar models with observations of eclipsing binaries \citep{1990ApJ...363L..33A} or open clusters \citep{1981A&A....93..136M} showed that this criterion actually underestimates the size of the convective core and proved the need for including extra-mixing processes in stellar models. If this fact is well-established, the nature of these processes (overshooting, rotationally induced mixing, internal gravity waves), the extent of the extra-mixed region, and their dependence on stellar parameters are still not fully understood. 

Another major uncertainty of stellar physics is the distribution and evolution of angular momentum \citep[e.g.][]{2013LNP...865....3M}. Rotation distorts stars and triggers supplementary hydrodynamical instabilities, which mixes stellar interiors and feeds back into rotation by transporting angular momentum. In particular, meridional circulation and shear instabilities are commonly accepted as operating mechanisms in stellar interiors. However, several studies \citep{1995ApJ...441..865C,1998A&A...334.1000M,2004A&A...425..229M} pointed out that these mechanisms are insufficient to reproduce the solar rotation profile as measured by helioseismology \citep[e.g.][]{1998ApJ...505..390S,2007Sci...316.1591G,2017A&A...604A..40F}. To explain the discrepancy between observations and rotating models, several additional processes of angular momentum transport have been proposed that either involves internal gravity waves \citep{2005Sci...309.2189C} or magnetic fields \citep{2005A&A...440L...9E}, but whether one is dominating over the others remains somewhat unclear. Similar uncertainties hold in evolved low-mass stars. The internal rotation rates of red giants, as measured by asteroseismology, is a few orders of magnitude smaller than what is predicted by rotating stellar models \citep{2012A&A...548A..10M}, indicating that one or several missing angular momentum transport processes are at work in these evolved stars \citep[e.g.][]{2014ApJ...796...17F,2015A&A...573A..80R,2015A&A...579A..31B,2017A&A...599A..18E}. 

Among the variety of pulsating stars, $\gamma$ Doradus ($\gamma$ Dor) and Slowly Pulsating B-type (SPB) stars are promising targets for obtaining constraints on both convective boundary mixing and angular momentum transport processes. These stars pulsate in high radial order gravity modes that probe the innermost radiative layers close to their convective core, where they have larger amplitude. Notably, \citet{2008MNRAS.386.1487M} and \citet{2013MNRAS.429.2500B} demonstrated that the properties of gravity modes (g modes) in these stars is particularly sensitive to the shape of the chemical gradient at the edge of the convective core. $\gamma$ Dor stars are late A- to early F-type main-sequence stars with masses between roughly 1.3 and 2.0 \msun. Their g-mode pulsations are thought to be excited by the modulation of the radiative flux at the base of their thin convective envelope, or so-called convective blocking mechanism \citep{2000ApJ...542L..57G,2005A&A...435..927D}. It is worth noting that a more specific interest of $\gamma$ Dors lies in the fact that they are progenitors of red giants. SPB stars are main-sequence stars that have spectral types B3-B9 and masses between 2.5 to 8 \msun. In these stars, the pulsations are driven by the $\kappa$-mechanism due to the metal opacity bump at $T \sim 2 \times 10^5$~K \citep{1993MNRAS.265..588D}. 

From the point of view of seismology, ground-based observations of these pulsators are especially impractical as their g-mode pulsations typically have periods of around one day. This seriously hindered the application of asteroseismology to $\gamma$ Dor and SPB stars until the advent of space missions dedicated to high precision photometry. In particular, \emph{Kepler} provided four years of highly precise and nearly uninterrupted photometry for many of these stars allowing the detection of g-mode series nearly equally spaced in periods as predicted by theory \citep{1980ApJS...43..469T,1951ApJ...114..373L}. Thus, in recent years, detailed seismic studies of slowly-rotating $\gamma$ Dor \citep{2014MNRAS.444..102K,2015MNRAS.447.3264S,2015MNRAS.454.1792K,2015EPJWC.10101005B,2016MNRAS.459.1201M} and SPB stars \citep{2014A&A...570A...8P,2015A&A...580A..27M} targeted by \emph{Kepler} could be undertaken. However, this regular structure does not hold for moderately and rapidly-rotating stars that, yet, constitutes the majority of these pulsating variables. Indeed, the projected equatorial velocities of B- and A-type stars is commonly around 100~km.s$^{-1}$ but, this can reach 250~km.s$^{-1}$ in some stars \citep{2002ApJ...573..359A,2007A&A...463..671R}.

Moderate or rapid rotation significantly affects the g-mode oscillation spectrum because of the Coriolis force. Observationally, the Fourier spectrum of the photometric time-series often contains hundreds of peaks with no apparent structure, at least not as predicted by classical perturbative treatments of rotation. This leads \citet{2015A&A...574A..17V} to develop a new algorithm to search for non-equally spaced g-mode series during the pre-whitening process. Successfully, they reported the detection of period spacing patterns in 50 $\gamma$ Dor stars \citep{2015ApJS..218...27V}. To identify the g modes and estimate their interior rotation rates, \citet{2016A&A...593A.120V} subsequently modelled the patterns following a direct approach. Using the asymptotic formulation of the traditional approximation that account for the main effects of the Coriolis force, the authors computed the gravity mode periods in a grid of stellar models representative of the $\gamma$ Dor instability strip and then fitted the observed patterns with a least-square minimisation. Following a similar approach, \citet{2017A&A...598A..74P} conducted a detailed seismic study of five rotating SPB stars in the \emph{Kepler} field.

Given the uncertainties contained in stellar models, we propose a different method that is model-independent and solely relies on the asymptotic traditional approximation. Section \ref{sec:method} revisits the theoretical background of the traditional approximation and introduces the details of this method. In Sect. \ref{sec:valid}, we validate it on synthetic spectra of representative $\gamma$ Dor models in which oscillations are computed with a 2D non-perturbative approach. We also assess the biases on our estimates of the buoyancy radius and near-core rotation frequency. In Sect. \ref{sec:kepler-targets}, we confirm the potential of such method on two already studied \textit{Kepler} targets, the $\gamma$ Dor star KIC12066947 and the SPB star KIC3459297. Finally, we discuss the limitations of the method in its current implementation and conclude in the last section.

%
%

\section{Method}
\label{sec:method}

\subsection{Theoretical background}
\label{subsec:th-back}
 
Formally, the pulsation periods are obtained by solving the equations of hydrodynamics perturbed by small fluctuations around an equilibrium structure.  Two approaches may be adopted: solving the whole system of equations numerically or conducting an asymptotic analysis assuming reasonable approximations. With this latter approach, in a non-rotating spherical star, \citet{1980ApJS...43..469T} demonstrated that the pulsation periods of high-order gravity modes $(\left|n\right| \gg \ell)$ can be well-approximated to first order by the following expression,
	\begin{equation}
	\label{eq:tassoul}
		P_{n,\ell,m} \approx \frac{P_0 \left(\left|n\right| + \epsilon \right)}{\sqrt{\ell\left(\ell+1\right)}},
	\end{equation}
where $n$ is the radial order, $\ell$ the angular degree and $m$ the azimuthal order of the pulsation mode. Note that, by convention, radial orders of g modes are negative. We adopt the convention that $m~>~0$ are prograde modes and $m~<~0$ retrograde modes. $\epsilon$ is near-constant and linked to the star's structure, and
	\begin{equation}
	\label{eq:p0}
		P_0 = 2\pi^2 \left(\int_{\mathcal{R}} \frac{N_\mathrm{BV}}{r} \mathrm{d}r\right)^{-1},
	\end{equation}
is the buoyancy radius. $N_\mathrm{BV}$ is the Brunt-Väisälä frequency and $\mathcal{R}$ represents the resonant cavity of g modes. Hence, modes of same degree $\ell$  and consecutive radial orders $n$ are equally spaced in period, which is characterised by constant period spacings defined as,
\begin{equation}
\Delta P_\mathrm{\ell} = P_{n+1,\ell,m} - P_{n,\ell,m} \approx \frac{P_0}{\sqrt{\ell\left(\ell+1\right)}}.
\end{equation}
This is a particularly interesting property of high-order g modes when searching for mode series in an observed spectrum. Note also that the pulsation periods does not depend on $m$.

	In a rotating star, pulsations are described by a coupled set of equations that is both computationally expensive and numerically complex to solve. First introduced in the context of Geophysics \citep{Eckart1960} and later applied to stellar pulsations \citep[e.g.][]{1987MNRAS.224..513L,1997ApJ...491..839L,2003MNRAS.340.1020T}, the traditional approximation of rotation (TAR) is a non-perturbative, but simplified, treatment of rotation that allows the separability of these equations while still acknowledging the main effects of the Coriolis force. Assuming the star is in solid-body rotation, the TAR neglects the horizontal component of the rotation vector. In other words, the radial motions due to the Coriolis force and the radial component of the Coriolis force associated with horizontal motions are discarded. This approximation is well-justified for low-frequency pulsations as it is discussed here \citep{1978A&A....70..597B}, except near the stellar centre where radial and horizontal motions become comparable. To obtain the separability of the system, the \citet{1941MNRAS.101..367C} approximation, which neglects the perturbation of the gravitational potential, is further made. Finally, the centrifugal distortion is neglected ensuring the spherical symmetry of the star's structure. This latter hypothesis can be justified by the fact that the g modes are mostly sensitive to the innermost radiative layers, where centrifugal distortion remains minimal.

Thus, in the reference frame in co-rotation with the star, the equations of pulsations become separable in the three spherical coordinates $(r,\theta,\varphi)$. The dependence in latitudinal coordinate is governed by the Hough functions, $\Theta_{\ell,m}$, that are the eigenfunctions of the well-known Laplace's tidal eigenvalue problem,
	\begin{equation}
	\label{eq:Laplace-tidal}
	\mathcal{L}_\mathrm{s}[\Theta_{\ell,m}\left(\mu,s\right)] = - \lambda_{\ell,m}\left(s\right)\Theta_{\ell,m}\left(\mu,s\right).
	\end{equation} 
where  $\mathcal{L}_\mathrm{s}$ is a linear operator, $\lambda_{\ell,m}$ are the eigenvalues of the problem and  $\mu = \cos \theta$. A complete derivation of the problem is derived, e.g., in \citet{1997ApJ...491..839L}. Both the Hough functions and associated eigenvalues depend on $\ell$, $m$, and the spin parameter $s = 2 P_{n,\ell,m}^\mathrm{co} / P_\mathrm{rot}$, where $P_{n,\ell,m}^\mathrm{co}$ is the mode period in the co-rotating frame and $P_\mathrm{rot}$ is the rotation period of the star. This latter reduced parameter appears naturally in the frame of the TAR, and is indicative of the influence of the Coriolis force on pulsation modes. The equation for the radial component is similar to that of the non-rotating case, except that $\ell(\ell+1)$ are replaced by the new eigenvalues $\lambda_{\ell,m}$. The azimuthal dependence stays identical.

As their form is akin to the non-rotating case, the asymptotic analysis of \citet{1980ApJS...43..469T} may also be applied to the pulsation equations obtained within the TAR. It follows a more general asymptotic formula for periods of high-order g modes in the co-rotating frame,
	\begin{equation}
	\label{eq:tar}
		P_{n,\ell,m}^\mathrm{co}\left(s\right) \approx \frac{P_0 \left( \left|n\right| + \epsilon \right)}{\sqrt{\lambda_{\ell,m}\left(s\right)}}.
	\end{equation}
The mode periods in the inertial frame, are subsequently deduced from,
	\begin{equation}
	\label{eq:co2in}
		P_{n,\ell,m}^\mathrm{in} = \frac{1}{\nu_{n,\ell,m}^\mathrm{co}+m\nurot}=  \frac{P_{{n,\ell,m}}^\mathrm{co}}{1 + m \nurot P_{n,\ell,m}^\mathrm{co}},
	\end{equation}
where $\nu_{n,\ell,m}^\mathrm{co}$ is the mode frequency in the co-rotating frame and $\nurot$ is the cyclic rotation frequency.
Therefore, in rotating stars, the degeneracy in azimuthal order is lifted and the regular structure of the spectrum no longer holds as period spacings now also depend on the spin parameter. 

\subsection{Behaviour of gravity modes in rotating stars}
\label{subsec:g-modes}

We recall here the behaviour of high-order g modes in rotating stars. We refer the reader to the works of \citet{2008MNRAS.386.1487M}, \citet{2013MNRAS.429.2500B} and \citet{2017MNRAS.465.2294O} for a more complete and detailed picture.

\begin{figure}
		\resizebox{\hsize}{!}{\includegraphics{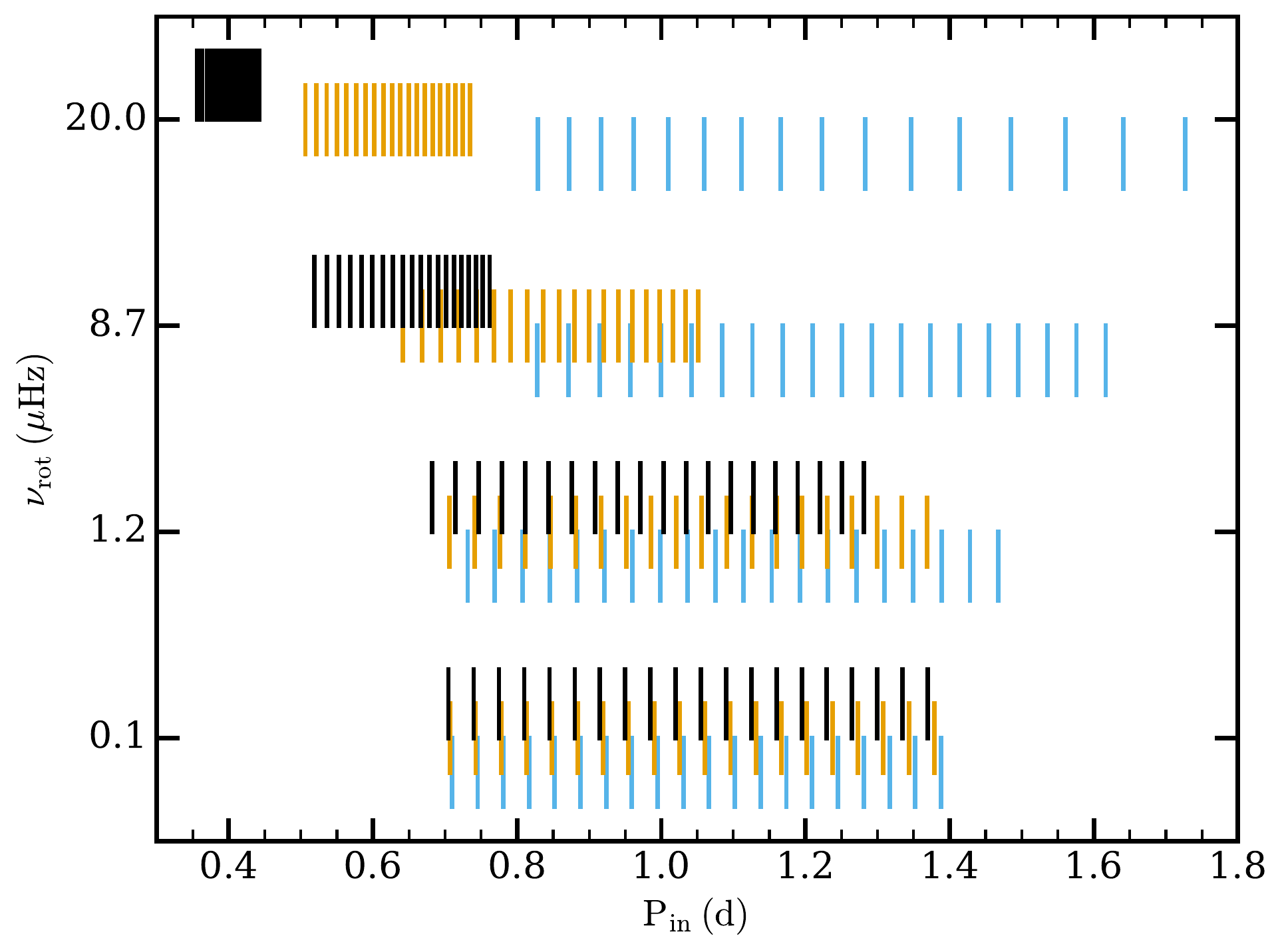}}
		\caption{Illustration of the rotational shift of observed pulsation periods, i.e. in the inertial frame, for four rotation rates. The oscillation spectra were computed within the asymptotic TAR (see Eq. \ref{eq:tar}) using $P_0 = 4320\: \mathrm{s}$, a value typical of $\gamma$ Dors. Black, orange and blue bars are dipole prograde ($m=1$), zonal ($m=0$) and retrograde ($m=-1$) modes, respectively. They are shifted vertically for clarity.}
		\label{fig:illustr-gmodes}
\end{figure}

 In slowly-rotating stars ($s \ll 1$), the spectrum is organised in well-separated multiplets $(n,\ell)$. Information on the rotation  of the star is carried by the rotational splittings, which are expressed by $\delta \nu_\mathrm{n,\ell} = \nu_{n,\ell,m} - \nu_\mathrm{n,\ell,0}$ in terms of the observed mode frequencies. At moderate rotation, when the rotation period approaches those of the modes ($s \sim 0.1$), these multiplets start to overlap and it becomes difficult to disentangle them visually. Nonetheless, period spacings remain almost constant, which makes the detection of a regular structure still possible \citep[see e.g.][]{2015EPJWC.10101005B}. For rapid rotation, when the rotation period is of the same order (or greater) than the pulsation periods, the structure of the spectrum is fundamentally different. Indeed, rotation tends to separate modes according to their geometry, i.e. to their values of $(\ell,m)$. In the inertial frame, prograde and zonal modes are shifted towards shorter periods whereas retrograde modes tend to drift towards longer periods. If the rotation frequency is sufficiently high, $(\ell,m)$ modes form clusters of modes distinct from each other. These clusters partially overlay otherwise. This evolution of the g-mode spectrum with rotation is illustrated on Fig.~\ref{fig:illustr-gmodes}.

Rotation also leaves its imprint on the period spacings. Indeed, in the inertial frame, the period spacings of prograde and zonal modes decrease almost linearly with increasing mode periods. At fixed rotation frequency, the slope of this linear trend is smaller for prograde than for zonal modes. Also, for a given star, this slope get steeper with increasing rotation rate. As for retrograde modes, the change of reference frame and rotation have competitive effects on the mode periods, and so, on the period spacings, resulting in more complicated behaviour. However, the spacings tend to have an upper trend in the $\Delta P_\mathrm{in}$-$P_\mathrm{in}$ plan for the most part.

As $\gamma$ Dor and SPB stars evolve on the main-sequence, a chemical composition gradient develops near the convective core boundary. If no mixing processes smooth this gradient, this leads to a local sharp variation of the Brunt-Väisälä frequency. Such buoyancy glitch periodically traps the g modes in the gradient region. As a result, the period spacings slightly oscillate around the linear trend. The period of this oscillatory component is related to the location of the glitch while the amplitude gives information about its abruptness \citep{2008MNRAS.386.1487M,2013MNRAS.429.2500B}.

\subsection{The method}
\label{subsec:method}

\begin{figure}
		\resizebox{\hsize}{!}{\includegraphics{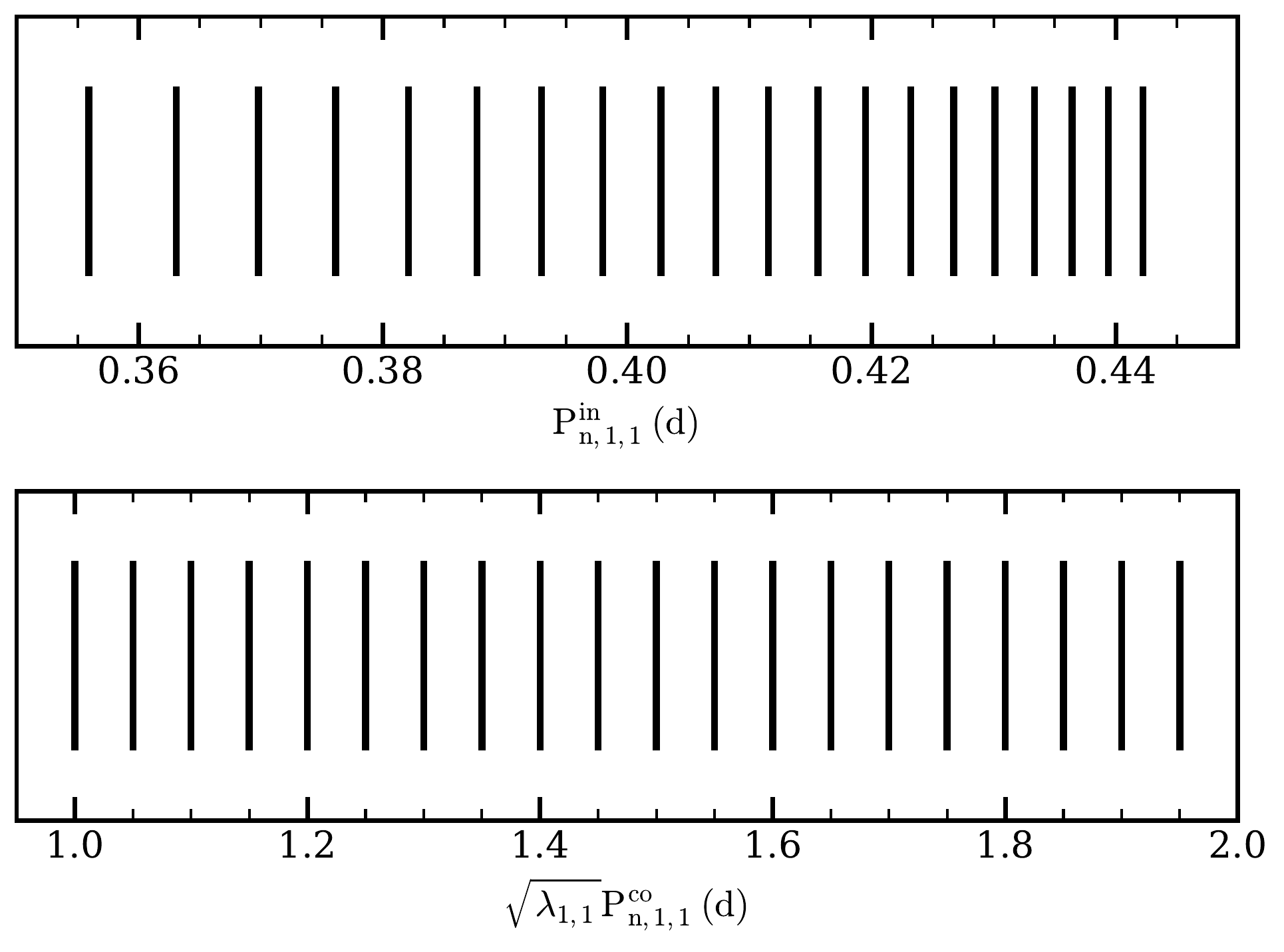}}
		\caption{Stretching of the pulsation periods in the example of a series of dipole prograde modes. Mode periods were generated using the asymptotic TAR with $\nurot = 20$~$\mu$Hz and $P_0 = 4320$ s (or 0.05~d). \textit{Top:} Before the stretching as they would be seen by an observer. \textit{Bottom:} After the change of reference frame and the stretching, assuming the correct mode identity and rotation rate.}
		\label{fig:example-atar}
	\end{figure}

In the frame of the asymptotic TAR, we highlight the possibility of recovering the equidistance of the period spacings by stretching the pulsation periods. Indeed, rearranging the terms of Eq. \ref{eq:tar},
 	\begin{equation}
	\label{eq:rearranged-tar}
		\sqrt{\lambda_{\ell,m}\left(s\right)} P_{n,\ell,m}^\mathrm{co} \approx P_0 \left( \left| n\right| + \epsilon \right),
	\end{equation}
shows that, by multiplying the period scale by the square root of the Laplace's eigenvalue $\lambda_{\ell,m}\left(s\right)$, where $s$ matches the star's rotation frequency, the associated $(\ell,m)$ modes become regularly spaced of $P_0$. This is illustrated on Fig. \ref{fig:example-atar} where we applied the stretching to a synthetic series of prograde dipole modes $(\ell~=~1,m~=~1)$.

On the basis of this property, we implemented a method to search for these new regularities in the g-mode spectrum of $\gamma$ Dor and SPB stars and, simultaneously infer $\nurot$ and $P_0$. Here, we describe the methodology used in detail. Similar stretching techniques have been developed to interpret the mixed mode oscillation patterns of red giant stars \citep{2015A&A...584A..50M}. 

First, while not absolutely necessary, we carry out the frequency analysis of the oscillation spectrum (e.g. by pre-whitening the periodogram). This simplifies greatly the subsequent steps and allows us to identify possible combination frequencies that arise from non-linear processes occurring in the star. 

The periodogram of $\gamma$ Dor and SPB stars is typically dense and often contains a large amount of frequency peaks, where  several series of $(\ell,m)$ modes may coexist. The equidistance of the period spacings can only be recovered for a given series of $(\ell,m)$ modes at once. In other words,  for the method to work correctly, the $(\ell,m)$ modes searched for need to be dominant in the list of frequencies used for the analysis. The strategy adopted here consists in selecting a part of the spectrum where we expect modes of given $(\ell,m)$ to be prevailing, relying on the behaviour of g modes in rotating stars (Sect.~\ref{subsec:g-modes}). In the current implementation, this step is performed by visual inspection of the periodogram or of the period distribution. Typically, we look for frequency groupings or mode density variations. Such filtering is obviously subjective and may be a matter of trial and error when the g-mode spectrum appears especially disordered. Automation of this step is possible and is currently under study. At this point, we consider that a given mode series is dominant over other series possibly present in the spectrum. 

Once the list of mode periods to analyse is established, we pick a guess for $\ell$ and $m$ and choose a range of rotation frequency to test. In practice, surface cancellation effects limits greatly the visibilities of high degree modes ($\ell \gtrsim 4$). As for the rotation rate, we restrain the interval to 0-35~$\mu$Hz, which largely contains all the rotation frequencies measured in $\gamma$ Dors and SPBs until now. Then, for each rotation rate, we compute the pulsation periods in the co-rotating frame using Eq.~\ref{eq:co2in} and stretch the spectrum according to Eq.~\ref{eq:rearranged-tar}. 

\begin{figure}
		\resizebox{\hsize}{!}{\includegraphics{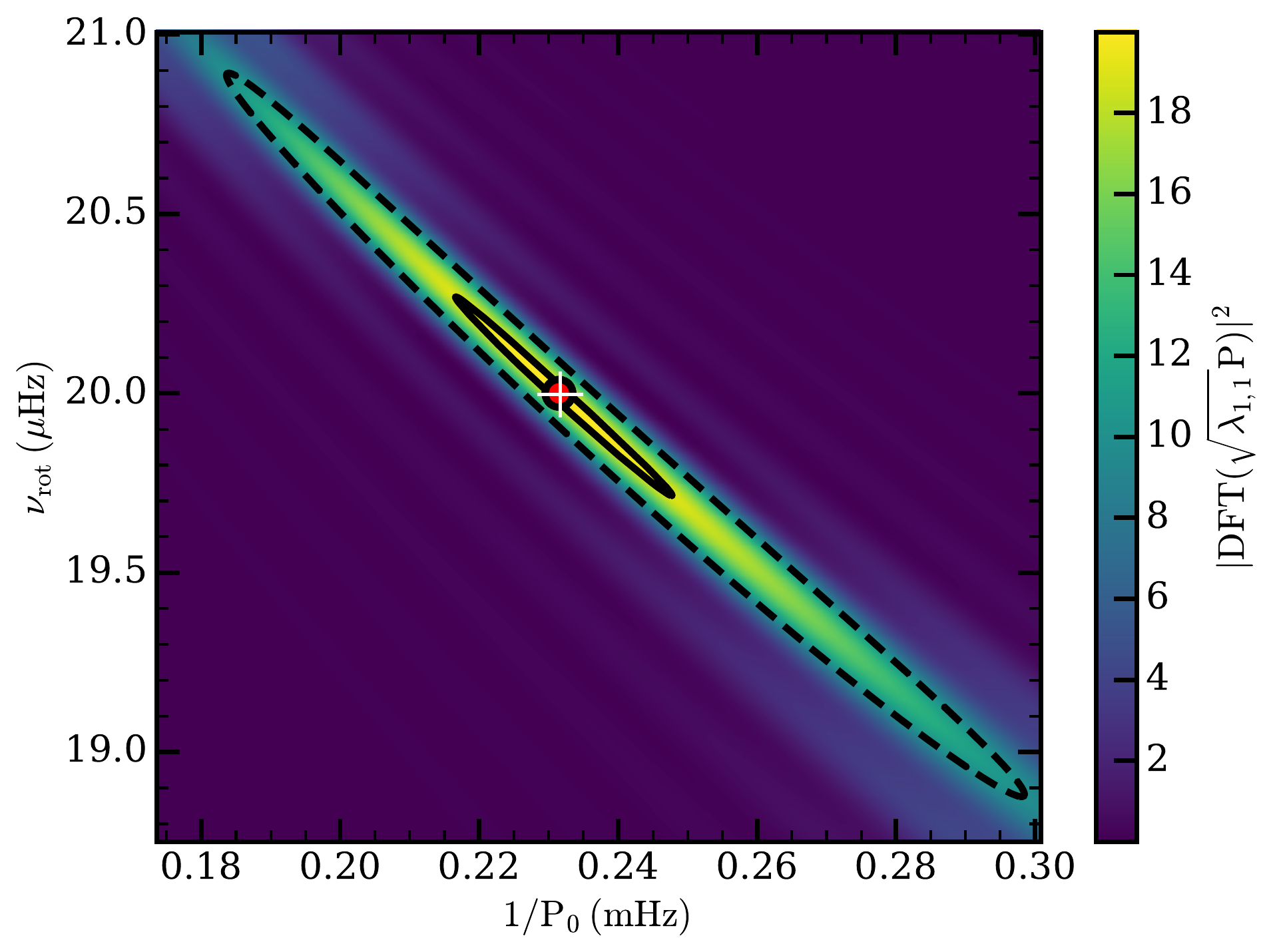}}
		\caption{DFT map obtained for the synthetic prograde mode series of Fig.~\ref{fig:example-atar}, as illustration. Red dot represents the true parameter values. White cross indicates the maximum of PSD. Solid and dashed lines are the contours at 95\% and 50\% of the maximum of PSD, respectively.}
		\label{fig:example-map}
	\end{figure}

To detect a regularity, we subsequently compute the power spectral density (PSD) of each stretched spectrum from the Discrete Fourier Transform (DFT). For a frequency $f$, this is given by,
\begin{equation}
\label{eq:dft}
\left| \mathrm{DFT}\left(\sqrt{\lambda_{\ell,m}}P_\mathrm{co}\right)\right|^2 \left(f\right)= \frac{1}{N} \left| \sum\limits_{i=1}^N  \mathrm{e}^{i 2 \pi f \sqrt{\lambda_{\ell,m}}P_\mathrm{i}^\mathrm{co}} \right|^2.
\end{equation}
where $P_\mathrm{i}^\mathrm{co}$ are the mode periods in the co-rotating frame and $N$ the total number of modes in the list. The DFT spectra obtained are then stacked vertically by increasing rotation rate to build a DFT map representative of the space of parameters explored. Figure \ref{fig:example-map} shows an example of such map. The detection of a regularity materialises into a characteristic ridge of high PSD. Such ridge is indicative of the correlation between $P_0$ and $\nurot$. Indeed, because $\nurot$ and $P_0$ are related through Eq.~\ref{eq:tar}, the effect of a small change in $\nurot$ can be nearly, but not exactly, counterbalanced by a small change in $P_0$. 

Finally, to ensure the detection is not coincidental, we compare the maximum of PSD to a threshold value computed from a false-alarm probabily $p = 0.01$ of having a peak generated by random noise. Further details are provided in Sect.~\ref{subsec:detection}. If greater than this threshold, the maximum of PSD is used as an estimator of $\nurot$ and $P_0$ (or more exactly $1/P_0$). The period \'echelle diagram of the stretched spectrum may then be plotted to have an overview of the modes actually participating in the regularity detected. Otherwise, the trial and error process may be continued by changing either the filtering of mode periods, the guess for $\ell$ and $m$, or the interval of rotation rate tested. \\

Briefly, the algorithm proceeds as follows,
\begin{enumerate}
			\item Establishing a list of frequencies (e.g. from Fourier analyses).	
			\item \textit{(Optional)} Filtering of pulsation modes according to their period values.
			\item Pick a guess for $(\ell,m)$ and choose the interval of rotation frequencies to test.
			\item For each rotational frequency:
			\begin{enumerate}
						\item Switch from the inertial to the co-rotating frame.
						\item Stretch the spectrum.
						\item Compute the DFT.
			\end{enumerate}
			\item Stack the DFT spectra obtained on top of another by increasing rotation rate (DFT map).
			\item Check if the maximum of PSD is significant.
			\begin{enumerate}
						\item If significant: formally identify modes and estimate $\nurot$ and $P_0$ from the maximum of PSD.
						\item If not: continue the trial and error process by returning to step 2 or 3.
			\end{enumerate}
\end{enumerate}

\subsection{Detection threshold}
\label{subsec:detection}

We derive here the detection threshold needed to achieve a specified probability of false alarm $p$ for a peak found in a DFT map. This $p$-value corresponds to the probability for that peak to be generated by pure noise. Each line of the map is computed as the norm square of the DFT of a stretched spectrum. From Eq.~\ref{eq:dft}, we can show that the noise in the power spectrum asymptotically (i.e. for large $N$) follows a two-degrees-of-freedom $\chi^2$ statistics  \citep[e.g.][]{AbramowitzStegun} when the period set $P_\mathrm{i}^\mathrm{co}$ is purely random\footnote{Since $P_\mathrm{i}^\mathrm{co}$ are random set, both real and imaginary parts of the DFT follow normal distributions according to the Central Limit theorem when $N$ is large enough. Hence, by definition, the norm square follows a two-degrees-of-freedom $\chi^2$ distribution.}. These statistics are correct as long as $N$, i.e. the number of periods in the set, is large enough. We have numerically verified that a dozen is sufficient in practice, that is often the case for real observations.
The normalisation used for the DFT in Eq.~\ref{eq:dft} ensures that the mean and variance of the statistics is 1. Thus, the probability for pure noise to generate, in a given DFT bin, a peak larger than a threshold $T$ is $p=\exp{-T}$. As a consequence, given $M$ independent bins in the DFT, the detection threshold is, 
\begin{equation}
 T = -\ln[1-(1-p)^{1/M}].
\end{equation}
The number of independent bins is then $M = \Delta f/\delta f$, where $\delta f$ is the spectral resolution and $\Delta f$ is the frequency window of interest, i.e. the range in which we look for $1/P_0$. The spectral resolution expresses $\delta f=(\sqrt{\lambda_{\ell,m}}P^\mathrm{co}_\mathrm{max} - \sqrt{\lambda_{\ell,m}}P^\mathrm{co}_\mathrm{min})^{-1}$ where $P^\mathrm{co}_\mathrm{max}$ and $P^\mathrm{co}_\mathrm{min}$ are the largest and smallest periods in the set.

For $p=0.01$ and a typical value of $M=50$, the detection threshold is then $T=8.5$.

\subsection{Estimating the uncertainties}
\label{subsec:uncertainties}

Due to the rotation-pulsation coupling, the mode identity, the rotation frequency and the buoyancy radius are intrinsically related which can give rise to degeneracies in the process of mode identification. Indeed, several mode identities can be found for a given $(\ell_1,m_1)$ mode series. This is due to the fact that, assuming another geometry $(\ell_2,m_2)$, it is sometimes possible to mimic the $(\ell_1,m_1)$ pattern by adjusting the values of $\nurot$ and $P_0$ accordingly. This problem has already been noticed by \citet{2016A&A...593A.120V}. To break the degeneracy, the authors compare the found value of the asymptotic period spacing $\Delta P_\ell$ (or equivalently $P_0$) to expected values in models, as a consistent check. We used the same strategy in this work. The buoyancy radius adopts a sufficiently narrow range of values in $\gamma$ Dor (roughly between 3500 and 5000 s) and SPB stars (approximatively from 5000 to 11500 s) to infer the correct mode identity in most cases \citep{2008MNRAS.386.1487M}. If that is insufficient, other expected properties of g modes can be considered such as the pulsation periods of excited modes or the slope of the pattern in the $\Delta P-P$ plan.\\

The rotation frequency and buoyancy radius as determined by our method is affected by two main sources of uncertainties: errors on oscillation mode periods and the adequacy of the asymptotic TAR to assess the properties of gravity modes. The former uncertainties can be taken into account by simply propagating the errors throughout the analysis. On the other hand, the evaluation of errors caused by the use of the asymptotic TAR is more complicated as it requires to know the "true" parameter values. It is nonetheless possible to assess the effect of a sharp feature of the stellar structure, such as a buoyancy glitch, or any other effect that does not affect the global trend of the period spacing patterns. This can be treated in the same way as a statistical error on the mode periods would be. Here, we describe the procedure used to evaluate the impact of these two types of uncertainties on the estimate of $\nurot$ and $P_0$. Other potential error sources or biases will be addressed in Sect. \ref{sec:valid}.

As a first step, it is necessary to quantify the relative contributions from each source. This is  achieved by performing a first analysis of the oscillation spectrum. From the values of $\nurot$ and $P_0$ obtained, we compute the mode periods in the asymptotic TAR using Eq. \ref{eq:tar} and identify each observed mode by comparing the two period values. If the standard deviation of the residuals $\left(P_{n,\ell,m} - P_{n,\ell,m}^\mathrm{TAR}\right)$  is significantly larger than the mean error on the observed mode periods, we consider it to be the dominant source of errors. Otherwise, we keep the individual uncertainty on each observed mode period that were determined during the mode extraction process.

Secondly, we propagate these errors by means of a Monte-Carlo simulation. To this end, we draw 500 random samples of the observed pulsation periods assuming the errors on them follow a normal distribution and are not correlated. In case the uncertainty on pulsation periods is not the dominant source of error, we adopt the standard deviation of the residual $\left(P_{n,\ell,m} - P_{n,\ell,m}^\mathrm{TAR}\right)$ as being the errorbar on all mode periods. Then we apply the method to each sample to get the distributions of $\nurot$ and $P_0$. The final values of the rotation rate and buoyancy radius are determined from the maximum of PSD during the first analysis and their errorbars are evaluated as the 1-$\sigma$ deviation of the distributions.

%
%

\section{Tests on synthetic spectra}
\label{sec:valid}

As validation, we tested the method on a set of synthetic oscillation spectra computed for representative models of $\gamma$ Doradus stars, which are presented in Sect. \ref{subsec:models-osc}.

In particular, we investigate the ability of the method to detect regularities in situations that are likely to be encountered in observations. This also lets us gauge potential biases introduced by the use of the asymptotic TAR as a prescription of the rotation-pulsation coupling. To this end,  we selected three test cases. Firstly, we examine the simple case of a uniformly rotating star with no structural glitch (Sect. \ref{subsec:rotation}). With the second model, we assess the method's robustness against the effect of a buoyancy glitch (Sect. \ref{subsec:buoglitch}). Finally, the effect of differential rotation is addressed with the last model (Sect. \ref{subsec:diff-rot}).

We restrained our study to dipole modes only as there are mostly those observed in high-order g-mode pulsators. For each synthetic spectrum, we analysed the three mode series independently ensuring that the maximum of PSD found is above the detection threshold defined in Sect \ref{subsec:detection}. The DFT maps obtained are compiled under the form of a contour map, where, for each of three DFT map, we depict the contours at 50\% and 95\% of the maximum of PSD. 

\subsection{Stellar structure and oscillation models}
\label{subsec:models-osc}

Here we explain the choices made for the calculation of the stellar structure and oscillation used to generate the synthetic oscillation spectra.
Let us first discuss the hypotheses made for the stellar structure calculation. Throughout this study, we opted for 1D spherical stellar models. Indeed, spherically symmetric 1D calculations have been tested against complete 2D calculations for g modes in polytropic models \citep{2012ASPC..462..389B}, as well as for a model of $\gamma$ Dor star \citep{2017MNRAS.465.2294O}. According to them, the 1D non-perturbative approach, which presents the advantage of requiring less numerical resources, gives satisfactory results compared to the full 2D approach.

Under this assumption, stellar models were computed with the stellar evolution code \textsc{cles} \citep{2008Ap&SS.316...83S} for  a mass of 1.86 M$_{\odot}$, and with initial helium mass fraction $Y = 0.27$ and metallicity $Z = 0.014$. We adopted the AGSS09 metal mixture \citep{2009ARA&A..47..481A} and corresponding opacity tables obtained with  OPAL opacities  \citep{1996ApJ...464..943I},  completed  at  low  temperature ($\log T <4.1$) with \citet{2005ApJ...623..585F} opacity tables. We  used  the  OPAL2001  equation  of  state  \citep{2002ApJ...576.1064R}  and  the  nuclear  reaction  rates  from  NACRE  compilation \citep{1999NuPhA.656....3A}, except  for  the $^{14}N(p,\gamma) ^{15}O$  nuclear  reaction, for which we adopted the cross-section from \citet{2004PhLB..591...61F}. Surface boundary conditions at $ T = T_{\rm eff}$ were provided by ATLAS model atmospheres \citep{1998HiA....11..646K}. Convection was treated using the mixing-length theory (MLT) formalism \citep{1958ZA.....46..108B}  with a parameter $\alpha_{\textrm{MLT}} = 1.70$.

We considered models with and without turbulent diffusion. Since the \textsc{cles} code does not include effects of rotation on transport of angular momentum or chemical species, we instead introduced mixing by turbulent diffusion, following the approach of \citet{2008MNRAS.386.1487M}. This reproduces an effect of rotationally-induced mixing that is quite similar to overshooting, but in addition tends to smooth chemical composition gradients inside the star. In models including this type of mixing, the coefficient of turbulent diffusion was set to $D_{\textrm{t}}=700$~$\mathrm{cm^{-2}\cdot s^{-1}}$ and kept constant to this value during evolution and in every layer of the models. This value was selected from a previous calibration to Geneva models with similar masses. Since the evolution code does not generate rotation profiles, those are then added ad hoc after evolution calculations. When a differential rotation profile is considered, it is adapted to the structure by the mean of an error function, centered right above the convective core, with a width depending on the profile of chemical gradient.

The oscillation modes of such 1D models were computed with a non-perturbative method using the \textsc{\textsc{acor}} oscillation code \citep{2012A&A...547A..75O,2015A&A...579A.116O}, which accounts for both the Coriolis and the centrifugal force. The \textsc{\textsc{acor}} code solves the hydrodynamics equations perturbed by Eulerian fluctuations, performing direct integration of the problem. The numerical method is based on a spectral multi-domain method which expands the angular dependence of eigenfunctions into spherical harmonics series, and whose radial treatment is particularly well adapted to the behaviour of equilibrium quantities in evolved models (at the interface of convective and radiative regions, and at the stellar surface).

In order to determine the range of radial orders to investigate,  we have relied on the non-adiabatic stability calculations provided in \cite{2013MNRAS.429.2500B}. All the pulsations spectra studied here are calculated for g-mode radial orders between $n=-50$ and $-20$.\\

\renewcommand{\arraystretch}{1.25}
\begin{table}[t]
    \begin{center}
        \caption[]
        {Properties of \textsc{\textsc{cles}} $\gamma$ Dor models used in Section \ref{sec:valid}.}
        \label{tab:modprop}
        \begin{tabular*}{\linewidth}{@{\extracolsep{\fill}} l c c }
        \hline
        \hline\noalign{\smallskip}
         											                                                            & Model A 		& Model B  		\\
        \noalign{\smallskip}\hline\noalign{\smallskip}
        $M/M_\odot$    					 												& 1.86 		         & 1.86  		    \\
        $T_\mathrm{eff}$   (K)			                                                   & 7960   			& 8025 		     \\
        $\log L/L_\odot$					                                                        & 1.269   			& 1.202 			\\
	   $\log g$								                                                        &	3.99				& 4.08				\\
	   $R/R_\odot$						                                                        &	2.27	      		& 2.07				\\
	   Age	(Myr)						                                                            &	1048             & 782				\\
	   $X_\mathrm{C}$																	&	0.34			    & 0.32	     		\\
	   $D_\mathrm{t}$ $\mathrm{(cm^{-2}\cdot s^{-1})}$			&	700			    & -						\\
	   $P_0$ (s)                                                                               			& 4579             & 4453      	     \\
        \noalign{\smallskip}\hline
        \end{tabular*}
    \end{center}
\end{table}

\subsection{Simple case: solid-body rotation and smooth period spacing patterns}
\label{subsec:rotation}

We started with the simplest case and computed a model of $\gamma$ Dor star in solid-body rotation with no glitches in the structure (Model A in Table \ref{tab:modprop}). Initially, we set the rotation frequency of the model at 7 $\mu$Hz, which is included in the range observed in $\gamma$ Dor stars \citep{2016A&A...593A.120V}. 

The top left pannel of Figure \ref{fig:rotall} shows the contours of the three DFT map obtained. While the three contours seem to agree on the values of $\nurot$ and $P_0$, there are still small departures from the true parameter values as measured directly from the model. Table \ref{tab:rotunif} compiles the results of the three analyses. We found that relative systematic errors on the parameters do not exceed a few percent but vary according to the mode geometry considered.  The stretched period \'echelle diagram for prograde dipole modes $(\ell = 1, m =1)$ is plotted on the top right panel of Fig. \ref{fig:rotall}, as an example. The ridge slightly weaves along a vertical line that would represent the ridge if the asymptotic TAR was perfectly suited. Period \'echelle diagrams for other series present identical features. Although model A has turbulent diffusion mixing, the chemical gradient at the convective core boundary is still quite substantial and gives rise to a light undulation of the period spacings.

In order to compare the systematic errors against other sources, we estimated the uncertainties on $\nurot$ and $P_0$ by proceeding as described in Sect. \ref{subsec:uncertainties} where we assumed a typical uncertainty on the mode periods of $1\times10^{-4}$~d. The standard deviation on the residuals $(P_{n,\ell,m} - P_{n,\ell,m}^\mathrm{TAR})$ are $6\times10^{-5}$, $1\times10^{-4}$ and $5\times10^{-4}$~d for the prograde, zonal and retrograde modes, respectively. The uncertainties derived in this manner stay one order of magnitude smaller than the biases highlighted in Table~\ref{tab:rotunif}, suggesting the latter is actually the critical source of errors on the parameter values. 

\begin{figure}
		\resizebox{\hsize}{!}{\includegraphics{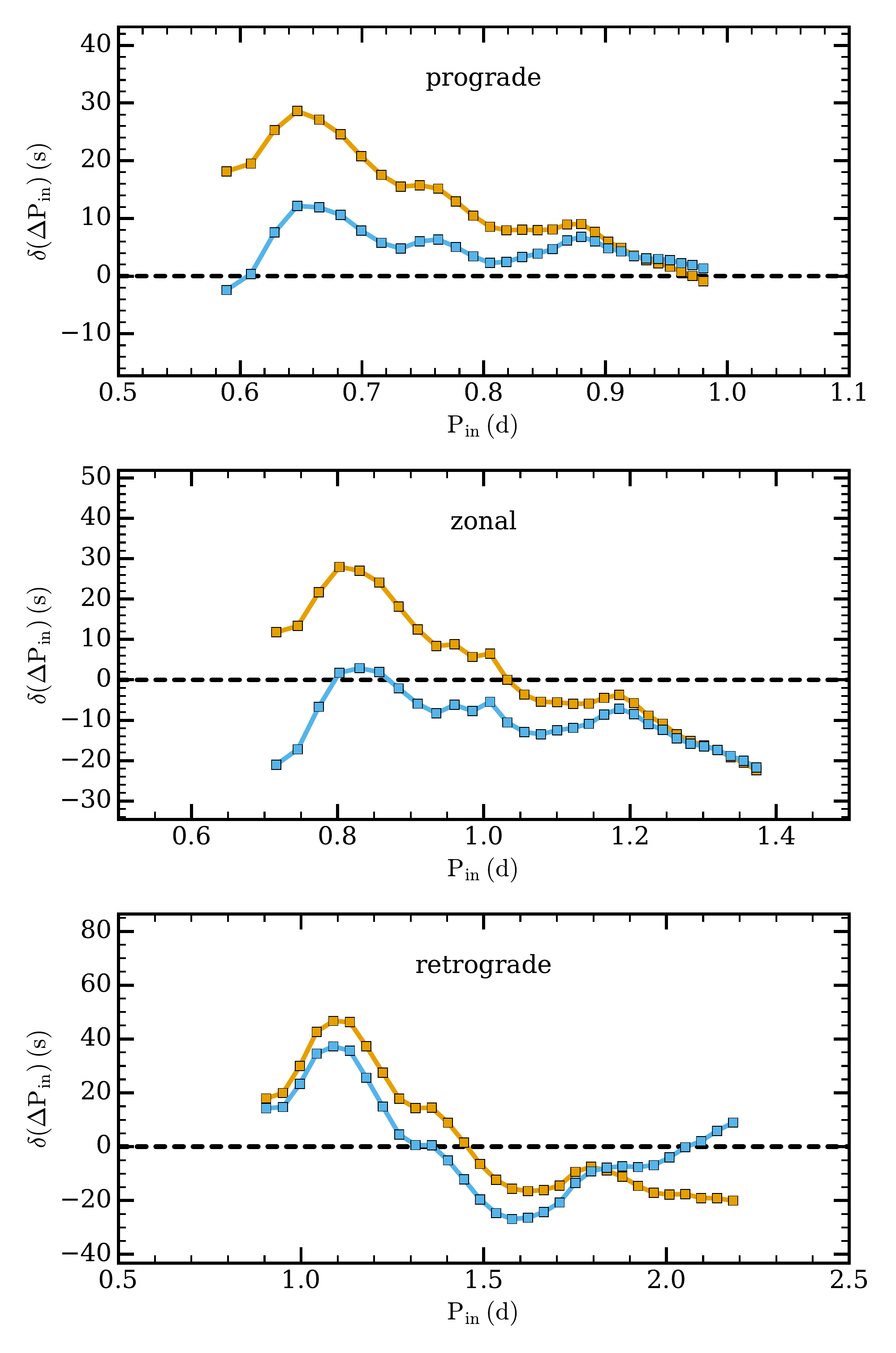}}
		\caption{Differences between the period spacings computed within the asymptotic TAR and those derived from complete calculations with \textsc{\textsc{acor}}. Orange squares: period spacings within the asymptotic TAR were computed using the parameter values of the model ($\nurot= 7$~$\mu$Hz and $P_0 = 4579$~s). Blue squares: using the parameter values found by the analysis of the spectrum (see Table \ref{tab:rotunif}).}
		\label{fig:tarvsmod2d7muHz}
\end{figure}

To investigate the provenance of these systematic errors, we compared the period spacings that can be obtained with the TAR using either the parameter values determined from the method analysis (Table \ref{tab:rotunif}) or from the model, to those of the complete calculations. From Eq. \ref{eq:tar}, the asymptotic period spacings within the TAR can be expressed as,
\begin{equation}
	\label{eq:period-spacing-tar}
	\Delta P_\mathrm{co} \simeq \frac{P_0}{\sqrt{\lambda_{\ell,m}\left(s\right)}\left(1 + \frac{1}{2} \frac{\mathrm{d}\ln \lambda_{\ell,m}\left(s\right)}{\mathrm{d} \ln s}\right)},
\end{equation} 
in the co-rotating frame \citep[see Appendix of][for a derivation]{2013MNRAS.429.2500B}. The spacings in the inertial frame are then derived from the following expression,
\begin{equation}
\label{eq:period-spacing-tar-frame-change}
	\Delta P_\mathrm{in} = \frac{\Delta P_\mathrm{co}}{\left(1+\frac{m}{2} s\left(n\right)\right)\left(1+\frac{m}{2} s\left(n+1\right)\right)}.
\end{equation}
These functions were interpolated on the period values of the complete calculations to work out the differences $\delta \left(\Delta P_\mathrm{in}\right) = \Delta P^\mathrm{TAR}_\mathrm{in} - \Delta P^\mathrm{\textsc{\textsc{acor}}}_\mathrm{in}$. Figure \ref{fig:tarvsmod2d7muHz} displays the results of each comparison. The parameter values outputted by the method are those that minimise the differences $\delta \left(\Delta P_\mathrm{in}\right)$, which turn out to be different from the true parameter values.  In other words, the stretched spectrum using the true values is less regular than if we used the output values of the method. This means that the asymptotic TAR does not perfectly model the spacings of complete calculations. 

The approximations made in the derivation of the asymptotic TAR, i.e. either the asymptotic approximation or the TAR (or both) then cause the disparity between the true parameter values and those recovered by the method. To investigate this further, we computed additional synthetic spectra. We set three rotation rates (7, 15 and 23~$\mu$Hz) in model A and computed the oscillation frequencies either within the non-asymptotic TAR or with \textsc{\textsc{acor}}. In this way, we are able to compare the relative contribution of each approximation to the bias and how it may vary with the rotation frequency. We limited this study to prograde modes ($\ell=1,m=1$) since other dipole modes are affected by an avoided crossing for model at 15 and 23~$\mu$Hz in the range of radial orders calculated with \textsc{acor}. Note also that prograde modes are mostly those detected in $\gamma$ Dor and SPB stars. The results of the analyses are listed in Table~\ref{tab:rotunif-bias}. The buoyancy radius is even more misestimated when the model is rotating fast, while no clear trend is visible for the recovered rotation frequencies. Moreover, the retrieved values are biased whether we take the non-asymptotic TAR or complete calculations. As could be expected, the errors obtained for the \textsc{acor} calculations are more important than for those in the non-asymptotic TAR. This indicates that both the asymptotic treatment and the TAR contribute to the bias. 

\renewcommand{\arraystretch}{1.25}
\begin{table}[t]
    \begin{center}
        \caption[]
        {Rotation frequencies and buoyancy radii as determined from our method for Model A in solid-body rotation (see Sect.  \ref{subsec:rotation}). Relative differences ($\delta \nurot$, $\delta P_0$) compared to the parameter inputs (7~$\mu$Hz, 4579~s) are also indicated.}
        \label{tab:rotunif}
        \begin{tabular*}{\linewidth}{@{\extracolsep{\fill}} l c c c c}
        \hline
        \hline\noalign{\smallskip}
        $(\ell,m)$			                    & $\nurot$ ($\mu$Hz) 		& $\delta\nurot$ (\%)	& $P_0$  (s) &	$\delta P_0$ (\%)\\
        \noalign{\smallskip}\hline\noalign{\smallskip}
	   $(1,1)$								&	6.95                               		&   0.75               				& 4495 		& 1.84\\
        $(1,0)$    	                            & 6.88   			                 		&   1.74              				& 4484   		& 2.09\\
        $(1, -1)$    				 		& 7.15                                 		&   2.14	       					& 4580   		&  0.02\\
        \noalign{\smallskip}\hline
        \end{tabular*}
    \end{center}
\end{table}

\renewcommand{\arraystretch}{1.25}
\begin{table*}[t]
    \begin{center}
        \caption[]
        {Recovered parameter values from the application of the method to synthetic oscillation spectra computed from Model A. The non-asymptotic TAR and the complete treatment of rotation (\textsc{acor}) are compared for three different rotation rates.}
        \label{tab:rotunif-bias}
        \begin{tabular*}{0.8\linewidth}{@{\extracolsep{\fill}} l c c c}
        \hline
        \hline\noalign{\smallskip}
          			     		& True values										 & \multicolumn{2}{c}{Recovered values}				\\
	               				& (Model A)											&  Non-asymptotic TAR 						& \textsc{acor} 			\\
        \noalign{\smallskip}\hline\noalign{\smallskip}
	    $\nurot$	($\mu$Hz)		&	7.00                               						&   6.97												& 6.95              \\
		$P_0$  (s)                           & 4579                                                       &  4531              									& 4495              \\
		\hline
	    $\nurot$	($\mu$Hz)		&	15.00                               				    &  14.95											& 14.94              \\
		$P_0$  (s)                           & 4579                                                       &  4498              									& 4459              \\
		\hline
	    $\nurot$	($\mu$Hz)		&	23.00                               				   &   22.95										    & 22.89\tablefootmark{1}              \\
		$P_0$  (s)                           & 4579                                                       &  4497             									& 4394\tablefootmark{1}             \\
        \noalign{\smallskip}\hline
        \end{tabular*}
    \end{center}
\tablefoot{
\tablefoottext{1}{For this specific model, an accumulation of higher $\ell$ modes occurs around the pulsation period of the radial order $n = - 45 $, which causes an avoided crossing.}
}
\end{table*}

\subsection{Effect of a buoyancy glitch}
\label{subsec:buoglitch}

The effect of a buoyancy glitch on the pulsation periods and therefore on the period spacings is not taken into account within the asymptotic TAR. Although the global trend of the period spacing pattern is not affected, it should be made clear if this impacts our estimates of $\nurot$ and $P_0$. Model B was computed without turbulent diffusion mixing so that a sharp chemical composition gradient grows at the convective core boundary forming a buoyancy glitch. We set the rotation frequency of this model to 7~$\mu$Hz to compute the mode periods.

Our results (Table \ref{tab:rotglitch} and middle left panel of Fig. \ref{fig:rotall}) show significant disagreements between the three dipole mode series. In addition, substantial deviations from the true parameter values are found. These are one order greater than for the smooth model used in Sect. \ref{subsec:rotation} reaching up to a maximum relative difference of $\sim$17~\% for $\nurot$ and  $\sim$6~\% for $P_0$.  The middle right panel of Fig. \ref{fig:rotall} displays the stretched period \'echelle diagram for prograde modes where the oscillatory component due to the buoyancy glitch is clearly visible. Other \'echelle diagrams show identical behaviour. 

In the general case, the oscillatory component, that is superposed to the smooth period spacing pattern in presence of a buoyancy glitch, is not symmetrical. By using the location of the maximum of PSD as an estimator of $P_0$, we actually find the mean spacing of the stretched pattern. The systematic errors underlined here can be, for the most part, attributed to the difference between this mean spacing and the true value of $P_0$. Note, however, that the biases found in Sect. \ref{subsec:rotation} also contribute here.

\renewcommand{\arraystretch}{1.25}
\begin{table}[t]
    \begin{center}
        \caption[]
        {Same as Table \ref{tab:rotunif} for Model B in solid-body rotation (see Sect. \ref{subsec:buoglitch}). Parameter inputs are 7~$\mu$Hz for $\nurot$ and 4453~s for $P_0$.}
        \label{tab:rotglitch}
        \begin{tabular*}{\linewidth}{@{\extracolsep{\fill}} l c c c c}
        \hline
        \hline\noalign{\smallskip}
        $(\ell,m)$			                    & $\nurot$ ($\mu$Hz) 		& $\delta\nurot$ (\%)	& $P_0$  (s) &	$\delta P_0$ (\%)\\
        \noalign{\smallskip}\hline\noalign{\smallskip}
	   $(1,1)$								&	6.67                               		&  4.71                				& 4233		     & 4.94\\
        $(1,0)$    	                            & 6.29   			                 		&  10.14            				& 4202   		&  5.64\\
        $(1, -1)$    				 		& 8.17                                 		&   16.77		        				    & 4645  		& 4.31\\
        \noalign{\smallskip}\hline
        \end{tabular*}
    \end{center}
\end{table}

\subsection{Differential rotation}
\label{subsec:diff-rot}

The TAR is a simplified treatment of rotation. One of its major and limiting hypotheses is the assumption of solid-body rotation, which is more invoked for the sake of mathematical simplifications rather than from physical considerations. As differentially rotating stars will almost surely be encountered in observational data, it is interesting to test if we are able to find regularities in the stretched spectrum for such stars and if so, investigate the impact of differential rotation on the estimate of the near-core rotation rate and buoyancy radius as obtained with our method. With this aim, we modelled a synthetic spectrum from Model A (Table \ref{tab:modprop}) on which we applied a two-zone rotation profile. The convective core rotates at $\nu_ \mathrm{rot, core} = 15 \: \mathrm{\mu Hz}$ while the rotation rate is $\nu_\mathrm{rot,env} = 7\: \mathrm{\mu Hz}$ in the envelope. In model A, the gradient of chemical composition at the convective core boundary is smoothed, which allows us to dismiss the effect of a buoyancy glitch hereafter.

\begin{figure*}
		\centering
		\includegraphics[width=18cm]{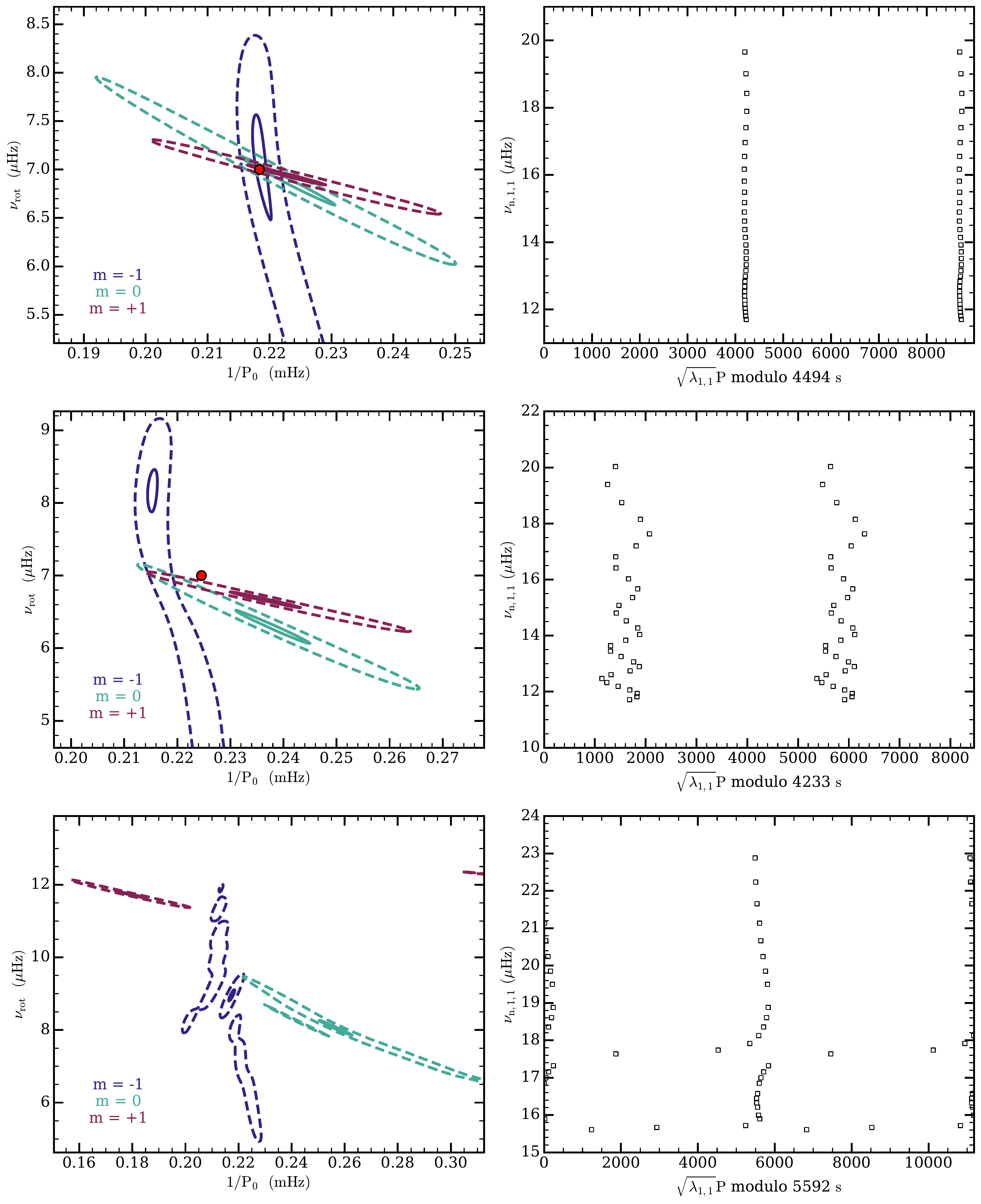}
		\caption{\textit{Left:} Contour maps at 95\% (solid) and 50\% (dashed) of the maximum of power spectral density  for Model A in solid-body rotation (\textit{top}), Model B in solid-body rotation (\textit{middle}) and Model A in differential rotation (\textit{bottom}). Colours are indicative of the type of modes $(\ell, m)$ on which was applied the method. Red dots indicates the input value of $\nurot$ and $1/P_0$ as measured from the models, when relevant. \textit{Right:} Example of stretched period \'echelle diagrams for prograde modes, which are plotted twice for clarity. }
		\label{fig:rotall}
\end{figure*}

The bottom left panel of Fig. \ref{fig:rotall} shows the contours for each DFT map. Table \ref{tab:rotdiff} lists the results of the analysis. Our results suggest that we should be in a position to find regularities in the stretched spectrum of a differentially rotating star. Moreover, the values of $\nurot$ and $P_0$ obtained  from each mode series differ from each other at a substantial level. These differences are superior to the expected biases and errors in an equivalent model in solid-body rotation (Model A and B, see Sect. \ref{subsec:rotation} and \ref{subsec:buoglitch}).  The bottom right panel of Fig. \ref{fig:rotall} represents the stretched period \'echelle diagram in the case of the prograde dipole modes. The ridge is almost vertical but also slightly curved. Moreover, two avoided crossings occur at $\sim$15.5 and $\sim$17.5~$\mu$Hz due to an accumulation of higher degree modes ($\ell=5$ and $\ell=7$, respectively) at these frequencies. This illustrates the extent of the TAR limitations. 

In a rotating star, the resonant cavity of gravity modes varies from a pulsation mode to another, which in turn impact on their properties like the pulsation period. While this variation is small (but still noticeable) within modes of same $(\ell,m)$, it can be quite large between modes of distinct $(\ell,m)$ couples.  Then, we have a signature of differential rotation. In particular, prograde modes probe deeper in the star than zonal and retrograde modes, which is consistent with much faster rotation. This is less clear for zonal and retrograde modes as the two rotation frequencies obtained are quite similar and this slight difference could indeed be explained by other factors. In particular, the corresponding ridges are not vertical in the stretched period \'echelle diagram. For this same reason, it is also difficult to reliably interpret the estimate of the buoyancy radii.

\renewcommand{\arraystretch}{1.25}
\begin{table}[t]
    \begin{center}
        \caption[]
        {Same as Table \ref{tab:rotunif} for Model A in differential rotation (see Sect. \ref{subsec:diff-rot}).}
        \label{tab:rotdiff}
        \begin{tabular*}{\linewidth}{@{\extracolsep{\fill}} l c c c}
        \hline
        \hline\noalign{\smallskip}
        $(\ell,m)$			                                                            & $\nurot$ ($\mu$Hz)		& $P_0$  (s)	\\
        \noalign{\smallskip}\hline\noalign{\smallskip}
        $(1, -1)$    				 												& 8.97 		         & 4598  \\
        $(1,0)$    	                                                                          & 8.06   			& 3899  \\
	   $(1,1)$																		&	11.74            &	5592\\
        \noalign{\smallskip}\hline
        \end{tabular*}
    \end{center}
\end{table}

%
%

\section{Applications on Kepler targets}
\label{sec:kepler-targets}

As a proof of concept, we  applied our method to two stars targeted by \textit{Kepler}, the $\gamma$ Dor star KIC12066947 and the SPB star KIC3459297. Both were previously studied in the literature hence giving us a point of comparison with other existing methods.

\subsection{Frequency analysis}
\label{subsec:mode-extraction}

For the two stars, the frequency analysis was performed using the classical iterative pre-whitening process, where, at each step, the peak with the highest amplitude in the periodogram was subtracted from the light curve. The statistical significance of each peak was derived on the basis of the false alarm probability  \citep{1982ApJ...263..835S}, which gives robust results for \textit{Kepler} data. To make sure that the peak extracted was not introduced during the pre-whitening process, we compare the amplitude of the extracted peak with value in the original data. If these deviate more than 25\% we disregard this peak. A similar procedure was also used by \citet{2015A&A...574A..17V}. A detailed description of the code used here will be given in Antoci et al. (in prep.).

Despite these precautions, noise peaks or spurious frequencies created by the pre-whitening process may still be extracted. Robustness of the stretching method is ensured as long as these frequencies remain in minority compared to the modes that are looked for, which justifies to be conservative in the frequency analysis.  On the contrary, actual pulsation mode frequencies of a series may not be extracted because of overly cautious criteria. This would introduce windowing in the DFT map or, in the worst case, prevent us from identifying the mode series. In this case, accepting frequencies with a worst agreement between extracted and original amplitude may improve the detection of the regularity. We tested the effect of missing orders in a modes series with simulated spectra. We found that a regularity could be detected as long as more than approximately half of the modes were present in the frequency list.

Line-of-sight motions of stars introduce Doppler shifts of the pulsation frequencies, which can be larger than their associated uncertainties in {\it Kepler} data \citep{2014MNRAS.445L..94D}. Here, we neglected these Doppler frequency shifts as we could not find line-of-sight velocity measurements in the literature for these two stars. The impact on the measured $\nurot$ and $P_0$ is expected to be contained within our internal uncertainties, even assuming large radial velocities ($\pm 150\: \mathrm{km.s^{-1}})$.

\subsection{The $\gamma$ Dor star KIC12066947}
\label{subsec:gammador}

\begin{figure*}
		\centering
		\includegraphics[width=17cm]{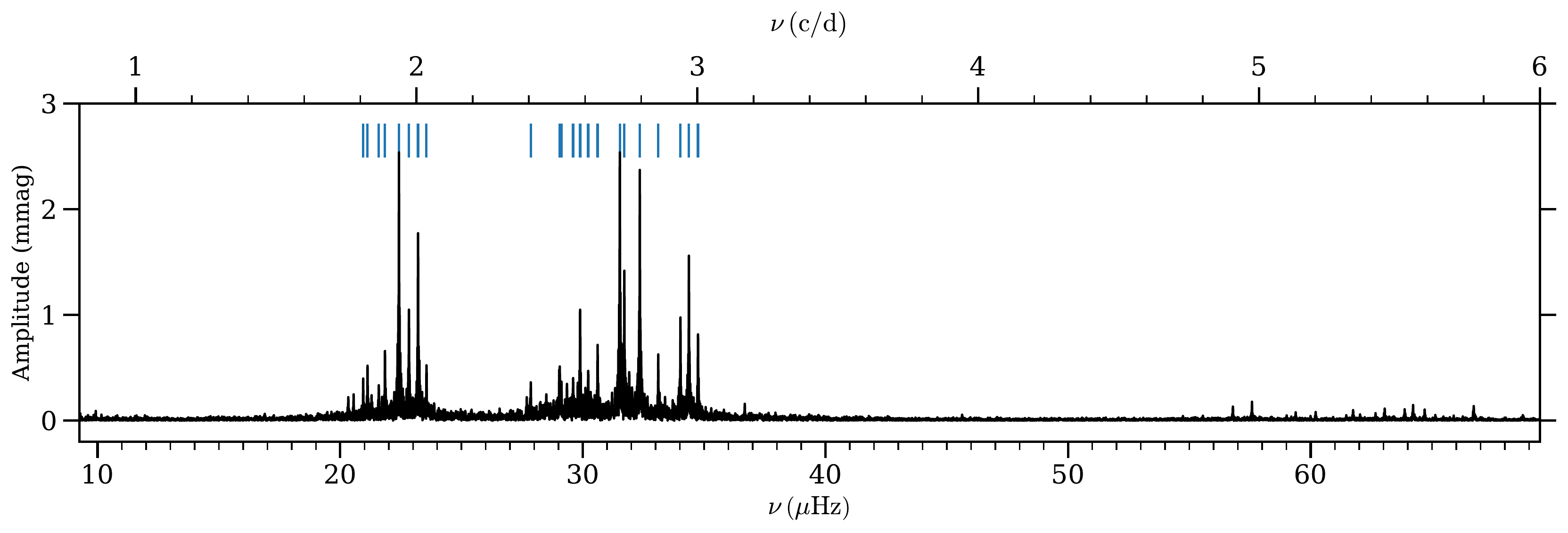}
		\caption{Part of the Lomb-Scargle periodogram of KIC12066947 computed from the \textit{Kepler} light curve. {\bf Blue bars} indicate the peak frequencies that were extracted and kept for the analysis.}
		\label{fig:GD-spectrum}
	\end{figure*}

The $\gamma$ Dor star  KIC12066947 was observed by the \textit{Kepler} spacecraft in long-cadence mode during quarters 0-1 and 10-17 thus gathering almost 670 days of high-quality photometric data. We could extract 22 peak frequencies from the Lomb-Scargle periodogram (Table \ref{tab:gdor-list}). As shown in Fig. \ref{fig:GD-spectrum}, most extracted peaks cluster in two groups in frequency roughly between 20 and $23\:  \mathrm{\mu Hz}$ (hereafter referred as cluster A),  and between 27 and $35\: \mathrm{\mu Hz}$ (cluster B). We analysed each cluster of peaks independently ensuring that we explored a sufficiently broad parameter space. 

\begin{figure*}
		\centering
		\includegraphics[width=17cm]{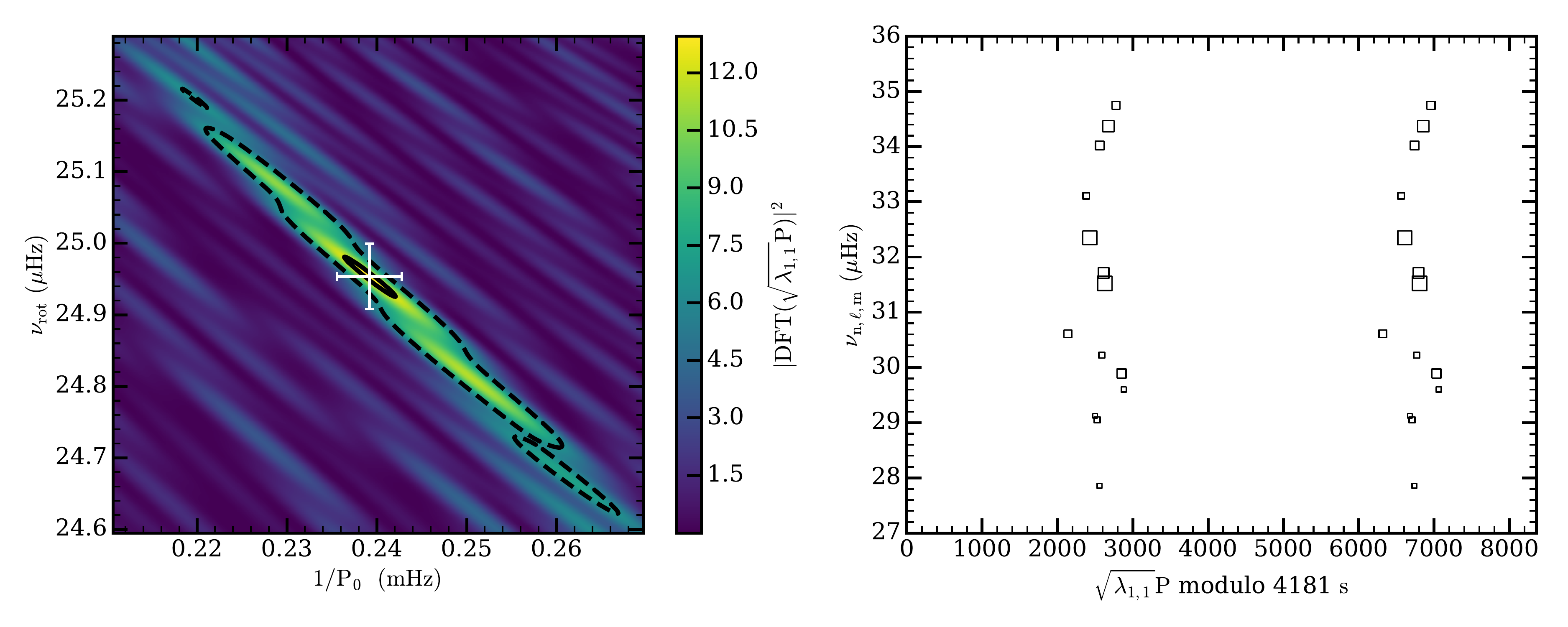}
		\caption{\textit{Left:}   DFT map resulting from the analysis of cluster B peaks, assuming they are prograde dipole modes $\left(\ell = 1, m =1\right)$. White cross is located at the maximum of power spectral density. Error bars were calculated as explained in Sect. \ref{subsec:uncertainties} and does not include systematic errors. Solid and dashed lines are the contours at 95\% and 50\% of the maximum of PSD.  \textit{Right:} Period \'echelle diagram  of the stretched spectrum composed of cluster B pulsation modes, plotted twice for clarity. Marker size is representative of their amplitude in the Lomb-Scargle periodogram.}
		\label{fig:clusterB-id}
	\end{figure*}

The analysis of cluster B frequencies leads to a detection for several trial values of $(\ell,m)$. However, unless we suppose they are (1,1) modes, the buoyancy radius given by the maximum of PSD is much larger than generally expected values for $\gamma$ Dor stars. We therefore identify these peaks as prograde dipole modes. The left panel of Fig. \ref{fig:clusterB-id} shows the DFT map obtained in this case. We found $\nurot = 24.95 \pm 0.05 \: \mathrm{\mu Hz}$ ($2.156 \pm 0.004 \: \mathrm{c/d} $) and $P_0  = 4181 \pm 63 \:\mathrm{s}$ ($0.0484 \pm 0.0008\: \mathrm{d}$). The quoted uncertainties were computed from a Monte-Carlo simulation as explained in Sect.~\ref{subsec:uncertainties} and does not account for systematic errors. The stretched \'echelle diagram built using these values is shown on the right panel of Fig.~\ref{fig:clusterB-id}. The ridge shows important curvatures indicating a substantial deviation from the asymptotic TAR. Such feature is however difficult to explain without a proper detailed study of the star. Furthermore, there is a great gap between the frequency at $\sim$ 27.8$~\mu$Hz and the rest of the ridge, causing a severe window effect in the DFT map. We checked if this has any influence on the derived values of $\nurot$ and $P_0$ by discarding the frequency and re-analysing this part of the spectrum. In this manner, we derived  $\nurot = 24.94 \pm 0.04 \: \mathrm{\mu Hz}$ and $P_0  = 4157 \pm 51 \:\mathrm{s}$ in good agreement with the previously determined values.

Despite searching for modes up to $\ell = 4$ and trying rotation frequencies from 2 to 35 $\mathrm{\mu Hz}$, we could not identify cluster A peaks. Given their high amplitudes, these peaks are more likely to be dipole modes. Using Eq. (\ref{eq:tar}) and the parameter values derived from the analysis of cluster B peaks, we computed the pulsation periods of dipole retrograde and zonal modes within the asymptotic TAR and then checked if they were consistent with the range of periods observed. Unfortunately, they do not agree well suggesting that these peaks do not originate from gravito-inertial mode pulsations.

The oscillation spectrum shows an excess of power around 55-65 $\mathrm{\mu Hz}$ with peaks of very low amplitude that are not extracted. Although they are located in the expected frequency range for (2,2) modes, the majority of these peak frequencies are simple combinations ($\nu_i+\nu_j$) or first harmonics ($2~\nu_i$) of the highest amplitude peaks.

KIC12066947 was studied by \citet{2015ApJS..218...27V,2016A&A...593A.120V} as part of their sample. Using the method of \citet{2015A&A...574A..17V} to analyse the oscillation spectrum, they detected two non-equidistant period spacing patterns. The first mode series consists of pulsation modes with frequencies in the range 27-37~$\mathrm{\mu Hz}$ and a downwards slope in the $\Delta P$ - $P$ plane. Modelling this pattern, they identified them as $(1,1)$ pulsation modes and found $\nurot = 25.00 \pm 0.10 \: \mathrm{\mu Hz}$  $(2.160 \pm 0.008 \: \mathrm{c/d}$). From their best fit model, they could also estimate the asymptotic period spacing $\Delta P_\ell = 2950\pm70 \: \mathrm{s}$, which translates into $P_0 = 4171 \pm 99 \:\mathrm{s}$ $(0.0484\pm 0.0012\: \mathrm{d}$) using Eq. (\ref{eq:tassoul}). These results agree very well with our determination from cluster B peaks. 
The second pattern has an upward slope consisting of pulsation frequencies in the range 19-22 $\mathrm{\mu Hz}$, i.e. corresponding to our cluster A. The authors could not identify them as g modes. The upward slope in the $\Delta P - P$ diagram suggest their retrograde character however, assuming the found $\nurot$ is accurate, their pulsation periods are much shorter than what would be expected from retrograde g modes. On the other hand, the authors could show that the period spacing pattern is consistent with retrograde Rossby modes (or r modes) using the asymptotic expression of \citet{2003MNRAS.340.1020T}. This seems also supported by theory and numerical computations that predict the excitation of r modes in fast-rotating stars due to an interaction between toroidal motion and rotation  \citep{2003MNRAS.340.1020T,2014A&A...569A..18S,2018MNRAS.474.2774S}. Interestingly, this would also explain why we could not identify these modes with our method. The TAR also predicts the r-mode periods with the same expression as Eq.~\ref{eq:tar}. Our stretching method can then also be applied to these modes by providing the $\lambda$ functions. First attempts are encouraging.

\subsection{The SPB star KIC3459297}
\label{subsec:spb}

\begin{figure*}
		\centering
		\includegraphics[width=17cm]{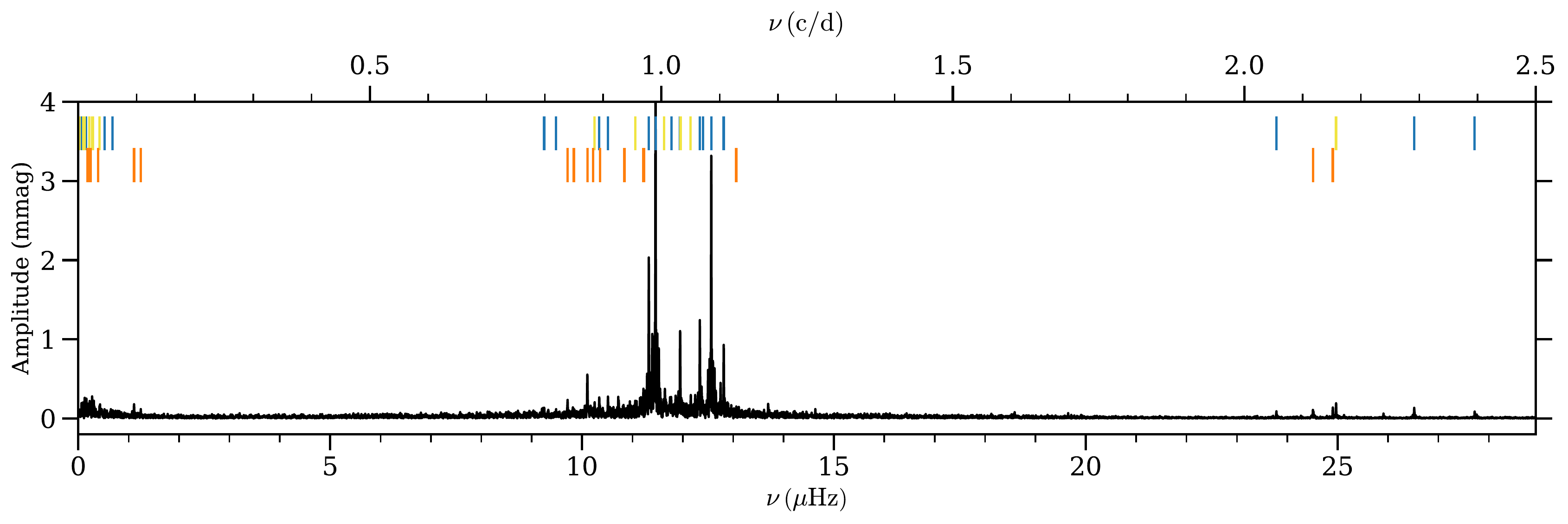}
		\caption{Lomb-Scargle periodogram of KIC3459297 computed from the \textit{Kepler} light curve. Blue bars indicate independent peak frequencies for which the amplitude extracted and that of the original data agrees within 25\%. Yellow bars are independent frequencies that are subsequently added to the frequency list (see text in Sect.~\ref{subsec:spb}). Orange bars are identified combination frequencies.}
		\label{fig:SPB-spectrum}
	\end{figure*}

The SPB star KIC3459297 was observed by \textit{Kepler} in long cadence mode during quarters Q0-5, 7-9, 11-13 and 15-17, which represents 1069 days of high-quality photometric data. The Lomb-Scargle periodogram computed from the light curve is plotted on Fig. \ref{fig:SPB-spectrum}.
We extracted 27 peak frequencies, of which 6 are found to be combinations and are discarded for the rest of the analysis (Table \ref{tab:spb-list}). Two groups of peaks clearly stand out. The cluster at very low frequency is most likely due to instrumental effects or may have been introduced by the detrending of the light curve. We analysed the other group in the 9-14~$\mathrm{\mu Hz}$ interval.

\begin{figure*}
		\centering
		\includegraphics[width=17cm]{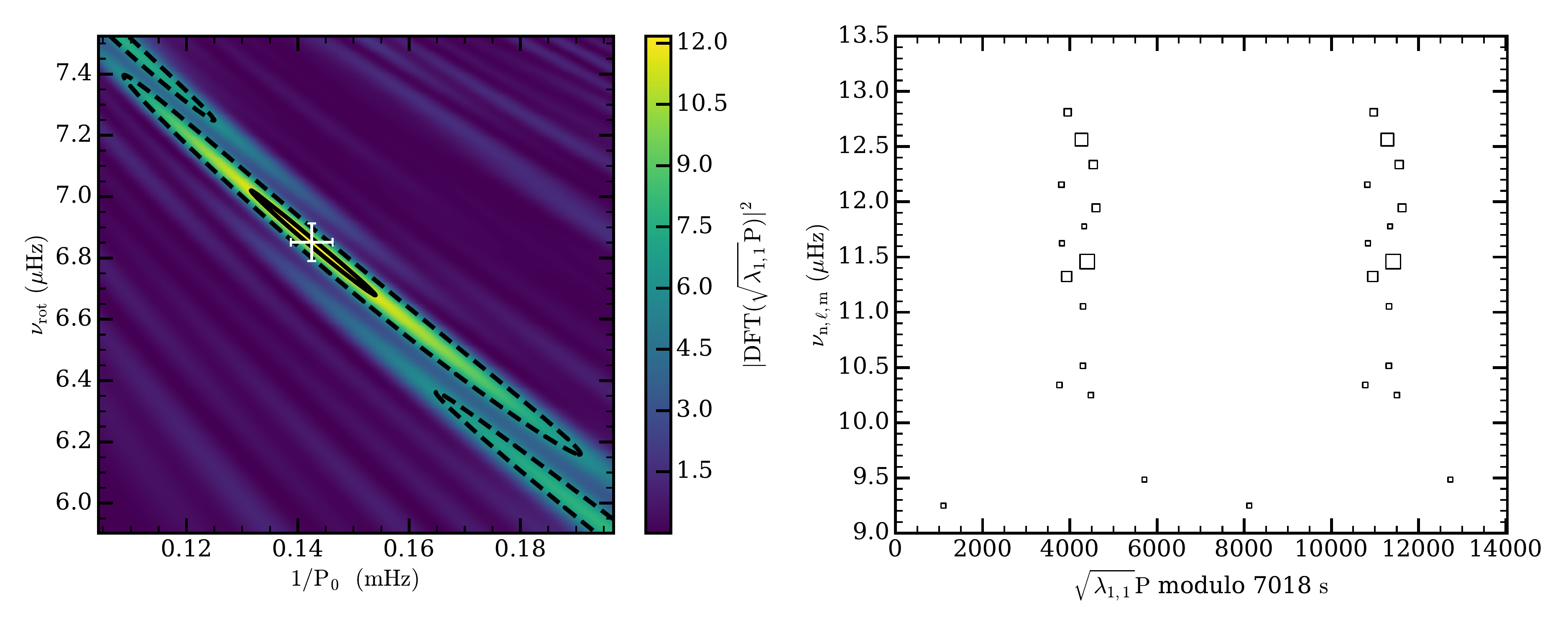}
		\caption{\textit{Left:}  DFT map resulting from the analysis of the group of peaks in the 9-14~$\mathrm{\mu Hz}$ interval of KIC3459297 periodogram, assuming they are prograde dipole modes $\left(\ell = 1, m =1\right)$. White cross is located at the maximum of PSD. Error bars were calculated as explained in Sect. \ref{subsec:uncertainties} and does not include systematic errors. Solid and dashed lines are the contours at 95\% and 50\% of the maximum of PSD. \textit{Right:} Period \'echelle diagram of the stretched spectrum, plotted twice for clarity. Marker size is representative of their amplitude in the Lomb-Scargle periodogram.}
		\label{fig:id-KIC3459297}
	\end{figure*}

From a first analysis, we could detect a series of modes, but due to numerous missing orders and consequent windowing, we could not reliably obtain $\nurot$ and $P_0$. In the following, we therefore included peaks with a lower agreement between the amplitude extracted and that of the original data (50\%). The~14 independent frequencies added are indicated by yellow bars on Fig.~\ref{fig:SPB-spectrum}. The detection threshold was reached for several trial $(\ell,m)$ values. However, similarly to the case of KIC12066947, only $(1,1)$ gives a buoyancy radius consistent with expected values for SPB stars. The peaks are thereby identified as prograde dipole modes. We estimated $\nurot = 6.85 \pm 0.07\:~\mathrm{\mu Hz}$ and $P_0 = 7018 \pm 190\:~\mathrm{s}$ (see left panel of Fig. \ref{fig:id-KIC3459297}). To obtain this estimate, we discarded the two isolated peaks at 9.25 and 9.38~$\mathrm{\mu Hz}$ to avoid the effect of windowing. Furthermore, as can be seen on the stretched period \'echelle diagram (right panel of Fig.~\ref{fig:id-KIC3459297}), the belonging of these two frequencies to the prograde series is debatable. Remarkably, the period spacing pattern shows an oscillatory behaviour, which is a typical signature of a buoyancy glitch \citep{2008MNRAS.386.1487M}.

The power excess between 23 and 28~$\mu$Hz is dominated by simple combination frequencies of high amplitude peaks ($\nu_i~ +~\nu_j$). The couple of remaining frequencies may be actual $(2,2)$ modes but cannot be formally identified with our method due to their restricted number.

This SPB star was studied in \citet{2017A&A...598A..74P}. The authors could find a long period spacing pattern consisting of 43 modes in the 9-14 $\mathrm{\mu Hz}$ domain with a downward slope in the $\Delta P - P$ plane. Following the method of \citet{2016A&A...593A.120V}, they identified the mode series as dipole prograde modes. Moreover, they estimated $\nurot= 7.3 \pm 0.5 \: \mathrm{\mu Hz}$ ($0.63 \pm 0.04\:\mathrm{c/d}$) and $\Delta P_\ell = 5840^{+950}_{-860} \: \mathrm{s}$, which translates into a buoyancy radius of $P_0 = 8260^{+1400}_{-1300} \: \mathrm{s}$ ($0.096^{+0.016}_{-0.014} \: \mathrm{d}$).  These findings are  compatible at a 1-$\sigma$ level with ours although somewhat greater. We made the choice to stay more conservative in the frequency analysis of the light curve. This results in a lesser number of extracted frequencies than \citet{2017A&A...598A..74P}.  We analysed the list of prograde mode periods provided in their Table 6 using our method. The results ($\nurot = 6.81\pm0.02\:~\mathrm{\mu Hz}$, $P_0 = 6857\pm66\:~\mathrm{s}$) are not significantly different from our first estimate showing that the difference does not come from the period list. 
\citet{2017A&A...598A..74P} found a second series of 8 modes in the 25-32~$\mathrm{\mu Hz}$ range that they do not identified as combination frequencies. They suggested these are (2,2) modes.

%
%

\section{ Discussion and conclusions}
\label{sec:conc}

We have developed and tested a stellar-model-independent method to disentangle the g-mode spectrum of moderately to rapidly-rotating $\gamma$ Dor and SPB stars. Using this method, we are able to simultaneously obtain the mode identity and estimate the near-core rotation frequency and buoyancy radius of these stars. Moreover, we successfully apply this method to two \textit{Kepler} targets. Our results are compatible with preceding determinations that used a different method \citep{2016A&A...593A.120V,2017A&A...598A..74P}, but the uncertainties on $\nurot$ and $P_0$ intrinsic to our method are substantially lower. It should be noted, however, that the dominant source of errors comes from the inadequacies of the asymptotic TAR on which \citet{2016A&A...593A.120V} also relied. It is also expected that the non-asymptotic formulation that was used by \citet{2017A&A...598A..74P} suffers from systematic biases.

While the method works well in most cases, precautions need to be taken during the analysis or in the interpretation of results in some specific cases.

Because we use the DFT to detect regularities in the stretched spectrum, modes in a given series must be in sufficient number so that a peak in the DFT map can be unambiguously attributed to a regularity, i.e. with an amplitude that is above the detection threshold (see Sect. \ref{subsec:detection}). We are also subject to the effects of windowing especially in case of missing orders in a ridge. Therefore, particular attention should be paid to the results when signatures of these effects are visible in the DFT map.

As a result of the non-linear coupling of pulsation modes, combination frequencies are common in the oscillation spectrum of $\gamma$ Dor and SPB stars. These combinations are generally considered to be harmonics or simple combinations of the frequencies of highest amplitude but this has been subject to debate as the underlying physical mechanism is not understood \citep[see][and references therein]{2015MNRAS.450.3015K}. Possible combination frequencies must be identified. Their analysis, as if there were independent frequencies, can lead to wrong interpretations. For instance, the first harmonics of $(1,1)$ may be misidentified as $(2,2)$ modes with the same rotation rate and buoyancy radius that can be derived from the fundamental frequencies. In this work, to avoid these misinterpretations, we searched for frequencies that could be explained as linear combinations ($\sum c_i \nu_i$ where $c_i \in \mathbb{Z}$) and we discarded those of order 3 or less ($\sum \left|c_i\right| \leq 3$) from the analyses.

$\gamma$ Dor and SPB stars are numerous in the \textit{Kepler} field of view. Few of them are slowly rotating and present rotational splittings that can be accurately interpreted with classical perturbative treatments of rotation. In this regard, we emphasise that the method presented in this paper complement the perturbative approach by allowing the interpretation of the spectrum of moderate to fast rotators. Thus, for rapid rotators, the analysis is straightforward as each mode series $(\ell,m)$ is well separated from the others. In moderate rotators, several g-mode series may overlap, which slightly complicates the identification process as the stretching function is relevant only for a given $(\ell,m)$ couple. This can be easily overcome by using the expected distribution of g modes in rotating stars as an indicator to guide the analysis (see Sect. \ref{subsec:g-modes}). As long as some $(\ell,m)$ modes are dominant in number on the portion of the spectrum studied, the associated regularity should be detectable. Then, proceeding iteratively,  i.e. identifying the dominant mode series, removing these modes from the list of frequencies, and re-analysing the remaining frequencies, allows the identification of overlapping series.

In the current implementation of the method, we assess the rotation-pulsation coupling in the framework of the asymptotic TAR. While the hypotheses behind this approximation are rather crude, it seems to offer reasonably accurate results. For a typical model of $\gamma$ Dor in uniform rotation and no structural glitch, we evaluated that the derived rotation rate and buoyancy radius are biased by only a few per cent. The TAR and the asymptotic approximations contribute to this bias in approximately equal proportions. 
In an equivalent model with a buoyancy glitch, the systematic errors can reach up to $\sim$17\% for $\nurot$ and ~6\% for $P_0$ because the effects of the buoyancy glitch on the pulsation periods is not accounted for in the asymptotic approximation. For the most observed dipole prograde modes, these biases remains limited to 5\% for both $\nurot$ and $P_0$. In a differentially rotating model, we show a clear signature of differential rotation although the full interpretation of the g-mode spectrum is not straightforward. The TAR assumes solid-body rotation and describes the pulsations in the co-rotating frame, which cannot be properly defined when there is differential rotation. Secondly, as mentioned before, the g-mode cavity changes from mode to mode. This questions the definition of a unique rotation frequency for a given mode series. 

We stress that these limitations and potential biases shown here are not specifically linked to our stretching method but are inherited from the asymptotic TAR, thus any method based on it would suffer the same issues. Furthermore, such method can be adapted to better prescriptions for pulsations in rotating stars. As long as the expression of pulsation periods adopts a suitable form, it is possible to define a stretching function such that the pulsation periods are regularly spaced once stretched. In which case, the determination of $\nurot$ and $P_0$ can be refined without modifying the core principle of the method. Recent theoretical studies \citep{2016A&A...587A.110P,2017A&A...598A.105P} have taken an interest in developing new and more accurate asymptotic theories on the basis of ray theory, which provides an interesting lead for future improvements of this method.

$\gamma$ Dor and SPB stars are present in great number in the wealth of data provided by \textit{Kepler}. Thanks to recent progress on the theory of pulsations in rotating stars as well as in the interpretation of their oscillation spectra, we are now in good position to  exploit the \textit{Kepler} data to their full potential. That promises to put tight constraints on convective-core boundary mixing and angular momentum transport.

\begin{acknowledgements} 
We acknowledge the International Space Science Institute (ISSI) for supporting the SoFAR international team.\footnote{\url{http://www.issi.unibe.ch/teams/sofar/}} The authors would like to thank M.J. Goupil for her comments and suggestions on the manuscript. J.B. acknowledges financial support from Programme National de Physique Stellaire~(PNPS) of CNRS/INSU, France, and from the Centre National d'\'Etudes Spatiales (CNES, France).  S.J.A.J.S. is funded by ARC grant for Concerted Research Actions, financed by the Wallonia-Brussels Federation.
 \end{acknowledgements}

\bibliographystyle{bibtex/aa}
\bibliography{bibtex/papers.bib}

\begin{thebibliography}{64}
\expandafter\ifx\csname natexlab\endcsname\relax\def\natexlab#1{#1}\fi

\bibitem[{{Abramowitz} \& {Stegun}(1964)}]{AbramowitzStegun}
{Abramowitz}, M. \& {Stegun}, I.~A. 1964, {Handbook of mathematical functions
  with formulas, graphs and mathematical tables} (Dover New York)

\bibitem[{{Abt} {et~al.}(2002){Abt}, {Levato}, \&
  {Grosso}}]{2002ApJ...573..359A}
{Abt}, H.~A., {Levato}, H., \& {Grosso}, M. 2002, \apj, 573, 359

\bibitem[{{Andersen} {et~al.}(1990){Andersen}, {Clausen}, \&
  {Nordstrom}}]{1990ApJ...363L..33A}
{Andersen}, J., {Clausen}, J.~V., \& {Nordstrom}, B. 1990, \apjl, 363, L33

\bibitem[{{Angulo} {et~al.}(1999){Angulo}, {Arnould}, {Rayet}, {Descouvemont},
  {Baye}, {Leclercq-Willain}, {Coc}, {Barhoumi}, {Aguer}, {Rolfs}, {Kunz},
  {Hammer}, {Mayer}, {Paradellis}, {Kossionides}, {Chronidou}, {Spyrou},
  {degl'Innocenti}, {Fiorentini}, {Ricci}, {Zavatarelli}, {Providencia},
  {Wolters}, {Soares}, {Grama}, {Rahighi}, {Shotter}, \& {Lamehi
  Rachti}}]{1999NuPhA.656....3A}
{Angulo}, C., {Arnould}, M., {Rayet}, M., {et~al.} 1999, Nuclear Physics A,
  656, 3

\bibitem[{{Asplund} {et~al.}(2009){Asplund}, {Grevesse}, {Sauval}, \&
  {Scott}}]{2009ARA&A..47..481A}
{Asplund}, M., {Grevesse}, N., {Sauval}, A.~J., \& {Scott}, P. 2009, \araa, 47,
  481

\bibitem[{{Ballot} {et~al.}(2012){Ballot}, {Ligni{\`e}res}, {Prat}, {Reese}, \&
  {Rieutord}}]{2012ASPC..462..389B}
{Ballot}, J., {Ligni{\`e}res}, F., {Prat}, V., {Reese}, D.~R., \& {Rieutord},
  M. 2012, in Astronomical Society of the Pacific Conference Series, Vol. 462,
  Progress in Solar/Stellar Physics with Helio- and Asteroseismology, ed.
  H.~{Shibahashi}, M.~{Takata}, \& A.~E. {Lynas-Gray}, 389

\bibitem[{{Bedding} {et~al.}(2015){Bedding}, {Murphy}, {Colman}, \&
  {Kurtz}}]{2015EPJWC.10101005B}
{Bedding}, T.~R., {Murphy}, S.~J., {Colman}, I.~L., \& {Kurtz}, D.~W. 2015, in
  European Physical Journal Web of Conferences, Vol. 101, European Physical
  Journal Web of Conferences, 01005

\bibitem[{{Belkacem} {et~al.}(2015){Belkacem}, {Marques}, {Goupil}, {Mosser},
  {Sonoi}, {Ouazzani}, {Dupret}, {Mathis}, \& {Grosjean}}]{2015A&A...579A..31B}
{Belkacem}, K., {Marques}, J.~P., {Goupil}, M.~J., {et~al.} 2015, \aap, 579,
  A31

\bibitem[{{Berthomieu} {et~al.}(1978){Berthomieu}, {Gonczi}, {Graff},
  {Provost}, \& {Rocca}}]{1978A&A....70..597B}
{Berthomieu}, G., {Gonczi}, G., {Graff}, P., {Provost}, J., \& {Rocca}, A.
  1978, \aap, 70, 597

\bibitem[{{B{\"o}hm-Vitense}(1958)}]{1958ZA.....46..108B}
{B{\"o}hm-Vitense}, E. 1958, \zap, 46, 108

\bibitem[{{Bouabid} {et~al.}(2013){Bouabid}, {Dupret}, {Salmon},
  {Montalb{\'a}n}, {Miglio}, \& {Noels}}]{2013MNRAS.429.2500B}
{Bouabid}, M.-P., {Dupret}, M.-A., {Salmon}, S., {et~al.} 2013, \mnras, 429,
  2500

\bibitem[{{Chaboyer} {et~al.}(1995){Chaboyer}, {Demarque}, \&
  {Pinsonneault}}]{1995ApJ...441..865C}
{Chaboyer}, B., {Demarque}, P., \& {Pinsonneault}, M.~H. 1995, \apj, 441, 865

\bibitem[{{Charbonnel} \& {Talon}(2005)}]{2005Sci...309.2189C}
{Charbonnel}, C. \& {Talon}, S. 2005, Science, 309, 2189

\bibitem[{{Cowling}(1941)}]{1941MNRAS.101..367C}
{Cowling}, T.~G. 1941, \mnras, 101, 367

\bibitem[{{Davies} {et~al.}(2014){Davies}, {Handberg}, {Miglio}, {Campante},
  {Chaplin}, \& {Elsworth}}]{2014MNRAS.445L..94D}
{Davies}, G.~R., {Handberg}, R., {Miglio}, A., {et~al.} 2014, \mnras, 445, L94

\bibitem[{{Dupret} {et~al.}(2005){Dupret}, {Grigahc{\`e}ne}, {Garrido},
  {Gabriel}, \& {Scuflaire}}]{2005A&A...435..927D}
{Dupret}, M.-A., {Grigahc{\`e}ne}, A., {Garrido}, R., {Gabriel}, M., \&
  {Scuflaire}, R. 2005, \aap, 435, 927

\bibitem[{{Dziembowski} {et~al.}(1993){Dziembowski}, {Moskalik}, \&
  {Pamyatnykh}}]{1993MNRAS.265..588D}
{Dziembowski}, W.~A., {Moskalik}, P., \& {Pamyatnykh}, A.~A. 1993, \mnras, 265,
  588

\bibitem[{{Eckart}(1960)}]{Eckart1960}
{Eckart}, C. 1960, Hydrodynamics of Oceans and Atmospheres (Pergamon Press,
  Oxford)

\bibitem[{{Eggenberger} {et~al.}(2017){Eggenberger}, {Lagarde}, {Miglio},
  {Montalb{\'a}n}, {Ekstr{\"o}m}, {Georgy}, {Meynet}, {Salmon}, {Ceillier},
  {Garc{\'{\i}}a}, {Mathis}, {Deheuvels}, {Maeder}, {den Hartogh}, \&
  {Hirschi}}]{2017A&A...599A..18E}
{Eggenberger}, P., {Lagarde}, N., {Miglio}, A., {et~al.} 2017, \aap, 599, A18

\bibitem[{{Eggenberger} {et~al.}(2005){Eggenberger}, {Maeder}, \&
  {Meynet}}]{2005A&A...440L...9E}
{Eggenberger}, P., {Maeder}, A., \& {Meynet}, G. 2005, \aap, 440, L9

\bibitem[{{Ferguson} {et~al.}(2005){Ferguson}, {Alexander}, {Allard}, {Barman},
  {Bodnarik}, {Hauschildt}, {Heffner-Wong}, \& {Tamanai}}]{2005ApJ...623..585F}
{Ferguson}, J.~W., {Alexander}, D.~R., {Allard}, F., {et~al.} 2005, \apj, 623,
  585

\bibitem[{{Formicola} {et~al.}(2004){Formicola}, {Imbriani}, {Costantini},
  {Angulo}, {Bemmerer}, {Bonetti}, {Broggini}, {Corvisiero}, {Cruz},
  {Descouvemont}, {F{\"u}l{\"o}p}, {Gervino}, {Guglielmetti}, {Gustavino},
  {Gy{\"u}rky}, {Jesus}, {Junker}, {Lemut}, {Menegazzo}, {Prati}, {Roca},
  {Rolfs}, {Romano}, {Rossi Alvarez}, {Sch{\"u}mann}, {Somorjai}, {Straniero},
  {Strieder}, {Terrasi}, {Trautvetter}, {Vomiero}, \&
  {Zavatarelli}}]{2004PhLB..591...61F}
{Formicola}, A., {Imbriani}, G., {Costantini}, H., {et~al.} 2004, Physics
  Letters B, 591, 61

\bibitem[{{Fossat} {et~al.}(2017){Fossat}, {Boumier}, {Corbard}, {Provost},
  {Salabert}, {Schmider}, {Gabriel}, {Grec}, {Renaud}, {Robillot},
  {Roca-Cort{\'e}s}, {Turck-Chi{\`e}ze}, {Ulrich}, \&
  {Lazrek}}]{2017A&A...604A..40F}
{Fossat}, E., {Boumier}, P., {Corbard}, T., {et~al.} 2017, \aap, 604, A40

\bibitem[{{Fuller} {et~al.}(2014){Fuller}, {Lecoanet}, {Cantiello}, \&
  {Brown}}]{2014ApJ...796...17F}
{Fuller}, J., {Lecoanet}, D., {Cantiello}, M., \& {Brown}, B. 2014, \apj, 796,
  17

\bibitem[{{Garc{\'{\i}}a} {et~al.}(2007){Garc{\'{\i}}a}, {Turck-Chi{\`e}ze},
  {Jim{\'e}nez-Reyes}, {Ballot}, {Pall{\'e}}, {Eff-Darwich}, {Mathur}, \&
  {Provost}}]{2007Sci...316.1591G}
{Garc{\'{\i}}a}, R.~A., {Turck-Chi{\`e}ze}, S., {Jim{\'e}nez-Reyes}, S.~J.,
  {et~al.} 2007, Science, 316, 1591

\bibitem[{{Guzik} {et~al.}(2000){Guzik}, {Kaye}, {Bradley}, {Cox}, \&
  {Neuforge}}]{2000ApJ...542L..57G}
{Guzik}, J.~A., {Kaye}, A.~B., {Bradley}, P.~A., {Cox}, A.~N., \& {Neuforge},
  C. 2000, \apj, 542, L57

\bibitem[{{Iglesias} \& {Rogers}(1996)}]{1996ApJ...464..943I}
{Iglesias}, C.~A. \& {Rogers}, F.~J. 1996, \apj, 464, 943

\bibitem[{{Keen} {et~al.}(2015){Keen}, {Bedding}, {Murphy}, {Schmid}, {Aerts},
  {Tkachenko}, {Ouazzani}, \& {Kurtz}}]{2015MNRAS.454.1792K}
{Keen}, M.~A., {Bedding}, T.~R., {Murphy}, S.~J., {et~al.} 2015, \mnras, 454,
  1792

\bibitem[{{Kurtz} {et~al.}(2014){Kurtz}, {Saio}, {Takata}, {Shibahashi},
  {Murphy}, \& {Sekii}}]{2014MNRAS.444..102K}
{Kurtz}, D.~W., {Saio}, H., {Takata}, M., {et~al.} 2014, \mnras, 444, 102

\bibitem[{{Kurtz} {et~al.}(2015){Kurtz}, {Shibahashi}, {Murphy}, {Bedding}, \&
  {Bowman}}]{2015MNRAS.450.3015K}
{Kurtz}, D.~W., {Shibahashi}, H., {Murphy}, S.~J., {Bedding}, T.~R., \&
  {Bowman}, D.~M. 2015, \mnras, 450, 3015

\bibitem[{{Kurucz}(1998)}]{1998HiA....11..646K}
{Kurucz}, R.~L. 1998, Highlights of Astronomy, 11, 646

\bibitem[{{Ledoux}(1951)}]{1951ApJ...114..373L}
{Ledoux}, P. 1951, \apj, 114, 373

\bibitem[{{Lee} \& {Saio}(1987)}]{1987MNRAS.224..513L}
{Lee}, U. \& {Saio}, H. 1987, \mnras, 224, 513

\bibitem[{{Lee} \& {Saio}(1997)}]{1997ApJ...491..839L}
{Lee}, U. \& {Saio}, H. 1997, \apj, 491, 839

\bibitem[{{Maeder} \& {Mermilliod}(1981)}]{1981A&A....93..136M}
{Maeder}, A. \& {Mermilliod}, J.~C. 1981, \aap, 93, 136

\bibitem[{{Maeder} \& {Zahn}(1998)}]{1998A&A...334.1000M}
{Maeder}, A. \& {Zahn}, J.-P. 1998, \aap, 334, 1000

\bibitem[{{Mathis} \& {Zahn}(2004)}]{2004A&A...425..229M}
{Mathis}, S. \& {Zahn}, J.-P. 2004, \aap, 425, 229

\bibitem[{{Meynet} {et~al.}(2013){Meynet}, {Ekstrom}, {Maeder}, {Eggenberger},
  {Saio}, {Chomienne}, \& {Haemmerl{\'e}}}]{2013LNP...865....3M}
{Meynet}, G., {Ekstrom}, S., {Maeder}, A., {et~al.} 2013, in Lecture Notes in
  Physics, Berlin Springer Verlag, Vol. 865, Lecture Notes in Physics, Berlin
  Springer Verlag, ed. M.~{Goupil}, K.~{Belkacem}, C.~{Neiner},
  F.~{Ligni{\`e}res}, \& J.~J. {Green}, 3

\bibitem[{{Miglio} {et~al.}(2008){Miglio}, {Montalb{\'a}n}, {Noels}, \&
  {Eggenberger}}]{2008MNRAS.386.1487M}
{Miglio}, A., {Montalb{\'a}n}, J., {Noels}, A., \& {Eggenberger}, P. 2008,
  \mnras, 386, 1487

\bibitem[{{Moravveji} {et~al.}(2015){Moravveji}, {Aerts}, {P{\'a}pics},
  {Triana}, \& {Vandoren}}]{2015A&A...580A..27M}
{Moravveji}, E., {Aerts}, C., {P{\'a}pics}, P.~I., {Triana}, S.~A., \&
  {Vandoren}, B. 2015, \aap, 580, A27

\bibitem[{{Mosser} {et~al.}(2012){Mosser}, {Goupil}, {Belkacem}, {Marques},
  {Beck}, {Bloemen}, {De Ridder}, {Barban}, {Deheuvels}, {Elsworth}, {Hekker},
  {Kallinger}, {Ouazzani}, {Pinsonneault}, {Samadi}, {Stello}, {Garc{\'{\i}}a},
  {Klaus}, {Li}, {Mathur}, \& {Morris}}]{2012A&A...548A..10M}
{Mosser}, B., {Goupil}, M.~J., {Belkacem}, K., {et~al.} 2012, \aap, 548, A10

\bibitem[{{Mosser} {et~al.}(2015){Mosser}, {Vrard}, {Belkacem}, {Deheuvels}, \&
  {Goupil}}]{2015A&A...584A..50M}
{Mosser}, B., {Vrard}, M., {Belkacem}, K., {Deheuvels}, S., \& {Goupil}, M.~J.
  2015, \aap, 584, A50

\bibitem[{{Murphy} {et~al.}(2016){Murphy}, {Fossati}, {Bedding}, {Saio},
  {Kurtz}, {Grassitelli}, \& {Wang}}]{2016MNRAS.459.1201M}
{Murphy}, S.~J., {Fossati}, L., {Bedding}, T.~R., {et~al.} 2016, \mnras, 459,
  1201

\bibitem[{{Ouazzani} {et~al.}(2012){Ouazzani}, {Dupret}, \&
  {Reese}}]{2012A&A...547A..75O}
{Ouazzani}, R.-M., {Dupret}, M.-A., \& {Reese}, D.~R. 2012, \aap, 547, A75

\bibitem[{{Ouazzani} {et~al.}(2015){Ouazzani}, {Roxburgh}, \&
  {Dupret}}]{2015A&A...579A.116O}
{Ouazzani}, R.-M., {Roxburgh}, I.~W., \& {Dupret}, M.-A. 2015, \aap, 579, A116

\bibitem[{{Ouazzani} {et~al.}(2017){Ouazzani}, {Salmon}, {Antoci}, {Bedding},
  {Murphy}, \& {Roxburgh}}]{2017MNRAS.465.2294O}
{Ouazzani}, R.-M., {Salmon}, S.~J.~A.~J., {Antoci}, V., {et~al.} 2017, \mnras,
  465, 2294

\bibitem[{{P{\'a}pics} {et~al.}(2014){P{\'a}pics}, {Moravveji}, {Aerts},
  {Tkachenko}, {Triana}, {Bloemen}, \& {Southworth}}]{2014A&A...570A...8P}
{P{\'a}pics}, P.~I., {Moravveji}, E., {Aerts}, C., {et~al.} 2014, \aap, 570, A8

\bibitem[{{P{\'a}pics} {et~al.}(2017){P{\'a}pics}, {Tkachenko}, {Van Reeth},
  {Aerts}, {Moravveji}, {Van de Sande}, {De Smedt}, {Bloemen}, {Southworth},
  {Debosscher}, {Niemczura}, \& {Gameiro}}]{2017A&A...598A..74P}
{P{\'a}pics}, P.~I., {Tkachenko}, A., {Van Reeth}, T., {et~al.} 2017, \aap,
  598, A74

\bibitem[{{Prat} {et~al.}(2016){Prat}, {Ligni{\`e}res}, \&
  {Ballot}}]{2016A&A...587A.110P}
{Prat}, V., {Ligni{\`e}res}, F., \& {Ballot}, J. 2016, \aap, 587, A110

\bibitem[{{Prat} {et~al.}(2017){Prat}, {Mathis}, {Ligni{\`e}res}, {Ballot}, \&
  {Culpin}}]{2017A&A...598A.105P}
{Prat}, V., {Mathis}, S., {Ligni{\`e}res}, F., {Ballot}, J., \& {Culpin}, P.-M.
  2017, \aap, 598, A105

\bibitem[{{Rogers} \& {Nayfonov}(2002)}]{2002ApJ...576.1064R}
{Rogers}, F.~J. \& {Nayfonov}, A. 2002, \apj, 576, 1064

\bibitem[{{Royer} {et~al.}(2007){Royer}, {Zorec}, \&
  {G{\'o}mez}}]{2007A&A...463..671R}
{Royer}, F., {Zorec}, J., \& {G{\'o}mez}, A.~E. 2007, \aap, 463, 671

\bibitem[{{R{\"u}diger} {et~al.}(2015){R{\"u}diger}, {Gellert}, {Spada}, \&
  {Tereshin}}]{2015A&A...573A..80R}
{R{\"u}diger}, G., {Gellert}, M., {Spada}, F., \& {Tereshin}, I. 2015, \aap,
  573, A80

\bibitem[{{Saio} {et~al.}(2018){Saio}, {Kurtz}, {Murphy}, {Antoci}, \&
  {Lee}}]{2018MNRAS.474.2774S}
{Saio}, H., {Kurtz}, D.~W., {Murphy}, S.~J., {Antoci}, V.~L., \& {Lee}, U.
  2018, \mnras, 474, 2774

\bibitem[{{Saio} {et~al.}(2015){Saio}, {Kurtz}, {Takata}, {Shibahashi},
  {Murphy}, {Sekii}, \& {Bedding}}]{2015MNRAS.447.3264S}
{Saio}, H., {Kurtz}, D.~W., {Takata}, M., {et~al.} 2015, \mnras, 447, 3264

\bibitem[{{Salmon} {et~al.}(2014){Salmon}, {Montalb{\'a}n}, {Reese}, {Dupret},
  \& {Eggenberger}}]{2014A&A...569A..18S}
{Salmon}, S.~J.~A.~J., {Montalb{\'a}n}, J., {Reese}, D.~R., {Dupret}, M.-A., \&
  {Eggenberger}, P. 2014, \aap, 569, A18

\bibitem[{{Scargle}(1982)}]{1982ApJ...263..835S}
{Scargle}, J.~D. 1982, \apj, 263, 835

\bibitem[{{Schou} {et~al.}(1998){Schou}, {Antia}, {Basu}, {Bogart}, {Bush},
  {Chitre}, {Christensen-Dalsgaard}, {Di Mauro}, {Dziembowski}, {Eff-Darwich},
  {Gough}, {Haber}, {Hoeksema}, {Howe}, {Korzennik}, {Kosovichev}, {Larsen},
  {Pijpers}, {Scherrer}, {Sekii}, {Tarbell}, {Title}, {Thompson}, \&
  {Toomre}}]{1998ApJ...505..390S}
{Schou}, J., {Antia}, H.~M., {Basu}, S., {et~al.} 1998, \apj, 505, 390

\bibitem[{{Scuflaire} {et~al.}(2008){Scuflaire}, {Th{\'e}ado}, {Montalb{\'a}n},
  {Miglio}, {Bourge}, {Godart}, {Thoul}, \& {Noels}}]{2008Ap&SS.316...83S}
{Scuflaire}, R., {Th{\'e}ado}, S., {Montalb{\'a}n}, J., {et~al.} 2008, \apss,
  316, 83

\bibitem[{{Tassoul}(1980)}]{1980ApJS...43..469T}
{Tassoul}, M. 1980, \apjs, 43, 469

\bibitem[{{Townsend}(2003)}]{2003MNRAS.340.1020T}
{Townsend}, R.~H.~D. 2003, \mnras, 340, 1020

\bibitem[{{Van Reeth} {et~al.}(2016){Van Reeth}, {Tkachenko}, \&
  {Aerts}}]{2016A&A...593A.120V}
{Van Reeth}, T., {Tkachenko}, A., \& {Aerts}, C. 2016, \aap, 593, A120

\bibitem[{{Van Reeth} {et~al.}(2015{\natexlab{a}}){Van Reeth}, {Tkachenko},
  {Aerts}, {P{\'a}pics}, {Degroote}, {Debosscher}, {Zwintz}, {Bloemen}, {De
  Smedt}, {Hrudkova}, {Raskin}, \& {Van Winckel}}]{2015A&A...574A..17V}
{Van Reeth}, T., {Tkachenko}, A., {Aerts}, C., {et~al.} 2015{\natexlab{a}},
  \aap, 574, A17

\bibitem[{{Van Reeth} {et~al.}(2015{\natexlab{b}}){Van Reeth}, {Tkachenko},
  {Aerts}, {P{\'a}pics}, {Triana}, {Zwintz}, {Degroote}, {Debosscher},
  {Bloemen}, {Schmid}, {De Smedt}, {Fremat}, {Fuentes}, {Homan}, {Hrudkova},
  {Karjalainen}, {Lombaert}, {Nemeth}, {{\O}stensen}, {Van De Steene}, {Vos},
  {Raskin}, \& {Van Winckel}}]{2015ApJS..218...27V}
{Van Reeth}, T., {Tkachenko}, A., {Aerts}, C., {et~al.} 2015{\natexlab{b}},
  \apjs, 218, 27

\end{thebibliography}


\begin{appendix}

\section{Lists of extracted frequencies}
\renewcommand{\arraystretch}{1.25}
	\begin{table}[h]
	\begin{center}	
	\caption[]
	{Frequencies for KIC12066947}
	\label{tab:gdor-list}
	\begin{tabular*}{0.7\linewidth}{@{\extracolsep{\fill}} l c c  }
	\hline
	\hline\noalign{\smallskip}
\#	  &  $\nu \: \mathrm{(\mu Hz)}$ & $\sigma_\nu \: \mathrm{(\mu Hz)}$   \\
	\noalign{\smallskip}\hline\noalign{\smallskip}
1	&	31.52532	&	0.00004		\\
2	&	22.42224	&	0.00003		\\
3	&	32.34848	&	0.00003		\\
4	&	23.20940	&	0.00006		\\
5	&	34.37178	&	0.00003		\\
6	&	31.71185	&	0.00006		\\
7	&	22.83401	&	0.00007	   \\
8	&	29.88961	&	0.00018		\\
9	&	34.02406	&	0.00004		\\
10	&	34.74865	&	0.00004	\\
11	&	30.61035	&	0.00020	\\
12	&	21.84573	&	0.00011	\\
13	&	33.10856	&	0.00010	\\
14	&	23.55624	&	0.00014	\\
15	&	30.22282	&	0.00036	\\
16	&	29.05031	&	0.00041	\\
17	&	21.12773	&	0.00016	\\
18	&	20.94863	&	0.00019	\\
19	&	29.60103	&	0.00049	\\
20	&	27.85636	&	0.00035	\\
21	&	29.12221	&	0.00057	\\
22	&	21.59091	&	0.00020	\\
\noalign{\smallskip}\hline
\end{tabular*}
\end{center}
\end{table}


\renewcommand{\arraystretch}{1.25}
	\begin{table}[]
	\begin{center}	
	\caption[]
	{Frequencies for KIC3459297}
	\label{tab:spb-list}
	\begin{tabular*}{0.85\linewidth}{@{\extracolsep{\fill}} l c c l }
	\hline
	\hline\noalign{\smallskip}
\#	  &  $\nu \: \mathrm{(\mu Hz)}$ & $\sigma_\nu \: \mathrm{(\mu Hz)}$ & \\
	\noalign{\smallskip}\hline\noalign{\smallskip}
1	&	11.45892	&	0.00002	&							 \\
2	&	12.56457	&	0.00001	&	     					 \\
3	&	11.32431	&	0.00003	&	     					 \\
4	&	12.33807	&	0.00003	&							 \\
5	&	11.94646	&	0.00004	&		 					\\
6	&	12.81174	&	0.00004	&							 \\
7	&	10.10531	&	0.00012	&	$ 2\nu_1-\nu_6$	 						\\
8	&	11.21814	&	0.00017	&	$\nu_1+\nu_2-\nu_6$	 			\\
9	&	12.15636	&	0.00013	&	 						\\
10	&	11.05385	&	0.00018	&						\\
11	&	0.27316	&	0.00037	&		 				\\
12	&	10.51394	&	0.00021	&	 					\\
13	&	0.12457	&	0.00039	&	 					\\
14	&	0.14769	&	0.00046	&						 \\
15	&	10.34129	&	0.00022	&						 \\
16	&	10.24943	&	0.00024	&	 					\\
17	&	0.10236	&	0.00057	&	 					\\
18	&	0.22518	&	0.00059	&	 $\nu_2-\nu_4$						\\
19	&	9.71168	&	0.00031	&	$\nu_3 - \nu_5 + \nu_{15}$	 \\
20	&	11.62550	&	0.00023	&						 \\
21	&	1.10575	&	0.00043	&	$\nu_2 - \nu_1$	 					\\
22	&	0.06223	&	0.00061	&						 \\
23	&	0.27982	&	0.00063	&						 \\
24	&	0.18418	&	0.00050	&	$\nu_4-\nu_9$	 					\\
25	&	10.35381	&	0.00031	&	$2\nu_1-\nu_2$						 \\
26	&	24.96855	&	0.00010	&						 \\
27	&	0.21483	&	0.00072	&		 				\\
28	&	10.21922	&	0.00039	&	$\nu_1-\nu_2+\nu_3$	 		\\
29	&	30.29326	&	0.00005	&						 \\
30	&	10.84022	&	0.00039	&	$\nu_1-\nu_2+\nu_5$, $\nu_1+\nu_3-\nu_5$	 \\
31	&	9.24711	&	0.00035	&	     				 \\
32	&	13.05932	&	0.00019	&	$\nu_2-\nu_1+\nu_5$			 \\
33	&	9.48398	&	0.00048	&	     				\\
34	&	11.23270	&	0.00038	&	$\nu_1-\nu_2+\nu_4$			 \\
35	&	0.41900	&	0.00089	&						 \\
36	&	0.28797	&	0.00089	&		 				\\
37	&	26.51890	&	0.00010	&		 				\\
38	&	1.24003	&	0.00065	&	$\nu_2-\nu_3$						 \\
39	&	11.77798	&	0.00038	&		 				\\
40	&	24.90285	&	0.00015	&	 $\nu_2+\nu_4$    					 \\
41	&	0.39570	&	0.00106	&	 $\nu_4-\nu_5$	 					\\
42	&	0.52098	&	0.00087	&						 \\
43	&	29.40322	&	0.00007	&						\\
44	&	24.50931	&	0.00016	&	$\nu_2+\nu_5$	 \\
45	&	23.78280	&	0.00015	&						 \\
46	&	0.68074	&	0.00101	&		 				\\
47	&	0.23975	&	0.00131	&	$\nu_6-\nu_2$	 \\
48	&	27.71991	&	0.00011	&						 \\
49	&	9.83608	&	0.00071	&	$2\nu_3-\nu_6$	 \\
50	&	0.02418	&	0.00155	&		 				\\
\noalign{\smallskip}\hline
\end{tabular*}
\end{center}
\end{table}

\end{appendix}

\end{document}